\documentstyle[aps,twocolumn,epsfig]{revtex}

\newcommand{\bq}{\begin{quotation}}
\newcommand{\eq}{\end{quotation}}

\newcommand{\veec}[4]{%
\begin{array}{c}{\!\!#1\!\!}\\{\!\!#2\!\!}\\{\!\!#3\!\!}\\{\!\!#4\!\!}\end{array}}

\begin{document}

\title{QBism, the Perimeter of Quantum Bayesianism}
\author{Christopher A. Fuchs}

\address{Perimeter Institute for Theoretical Physics\\
Waterloo, Ontario N2L 2Y5, Canada\medskip\\
{\tt cfuchs@perimeterinstitute.ca}}


\maketitle

\begin{abstract}
This article summarizes the Quantum Bayesian~\cite{Caves02,Fuchs02,Fuchs04,Caves07,Fuchs09a,Fuchs10,Timpson08} point of view of quantum mechanics, with special emphasis on the view's outer edges---dubbed QBism.\footnote{Quantum Bayesianism, as it is called in the literature, usually refers to a point of view on quantum states originally developed by C.~M. Caves, C.~A. Fuchs, and R.~Schack.  The present work, however, goes far beyond those statements in the metaphysical conclusions it draws---so much so that the author cannot comfortably attribute the thoughts herein to the triumvirate as a whole. Thus, the term QBism to mark some distinction from the known common ground of Quantum Bayesianism. Needless to say, the author takes sole responsibility for any inanities herein.}  QBism has its~roots in per\-sonal\-ist Bayesian probability theory, is crucially dependent upon the tools of quantum information theory, and most recently, has set out to investigate whether the physical world might be of a type sketched by some false-started philosophies of 100 years ago (pragmatism, pluralism, nonreductionism, and meliorism).  Beyond conceptual issues, work at Perimeter Institute is focussed on the hard technical problem of finding a {\it good\/} representation of quantum mechanics purely in terms of probabilities, without amplitudes or Hilbert-space operators.  The best candidate representation involves a mysterious entity called a symmetric informationally complete quantum measurement.  Contemplation of it gives a way of thinking of the Born Rule as an {\it addition\/} to the {\it rules\/} of probability theory, applicable when an agent considers gambling on the consequences of his interactions with a newly recognized universal capacity:  dimension (formerly Hilbert-space dimension).  (The word ``capacity'' should conjure up an image of something like gravitational mass---a body's mass measures its capacity to attract other bodies.  With hindsight one can say that the founders of quantum mechanics discovered another universal capacity, ``dimension.'')  The article ends by showing that the egocentric elements in QBism represent no impediment to pursuing quantum cosmology and outlining some directions for future work.
\end{abstract}

\section{A Feared Disease}

The start of the new decade has just passed and so has the media frenzy over the H1N1 flu pandemic.  Both are welcome events.  Yet, as misplaced as the latter turned out to be, it did serve to remind us of a basic truth:  That a healthy body can be stricken with a fatal disease which to outward appearances is nearly identical to a common yearly annoyance.  There are lessons here for quantum mechanics.  In the history of physics, there has never been a healthier body than quantum theory; no theory has ever been more all-encompassing or more powerful.  Its calculations are relevant at every scale of physical experience, from subnuclear particles, to table-top lasers, to the cores of neutron stars and even the first three minutes of the universe.  Yet since its founding days, many physicists have feared that quantum theory's common annoyance---the continuing feeling that something at the bottom of it does not make sense---may one day turn out to be the symptom of something fatal.

There is something about quantum theory that is different in character from any physical theory posed before.  To put a finger on it, the issue is this:  The basic statement of the theory---the one we have all learned from our textbooks---seems to rely on terms our intuitions balk at as having any place in a fundamental description of reality.  The notions of ``observer'' and ``measurement'' are taken as primitive, the very starting point of the theory.  This is an unsettling situation!  Shouldn't physics be talking about {\it what is\/} before it starts talking about {\it what will be seen\/} and who will see it?  Perhaps no one has put the point more forcefully than John Stewart Bell~\cite{Bell90}:
\begin{quote}\small
What exactly qualifies some physical systems to play the role of `measurer'?  Was the wavefunction of the world waiting to jump for thousands of millions of years until a single-celled living creature appeared?  Or did it have to wait a little longer, for some better qualified system \ldots\ with a PhD?
\end{quote}
One sometimes gets the feeling---and this is what unifies many a diverse quantum foundations researcher---that until this issue is settled, fundamental physical theory has no right to move on.  Worse yet, that to the extent it does move on, it does so only as the carrier of something insidious, something that will eventually cause the whole organism to stop in its tracks.  ``Dark matter and dark energy?  Might these be the first symptoms of  something systemic?  Might the problem be much deeper than getting our quantum fields wrong?'' --- This is the kind of fear at work here.

So the field of quantum foundations is not unfounded; it is absolutely vital to physics as a whole.  But what constitutes ``progress'' in quantum foundations?  How would one know progress if one saw it?  Through the years, it seems the most popular strategy has taken its cue (even if only subliminally) from the tenor of John Bell's quote:  The idea has been to remove the observer from the theory just as quickly as possible, and with surgical precision.  In practice this has generally meant to keep the {\it mathematical structure\/} of quantum theory as it stands (complex Hilbert spaces, operators, tensor products, etc.), but, by hook or crook, find a way to tell a story about the mathematical symbols that involves no observers at all.

In short, the strategy has been to reify or objectify all the mathematical symbols of the theory and then explore whatever comes of the move.  Three examples suffice to give a feel:  In the de~Broglie~--~Bohm ``pilot wave'' version of quantum theory, there are no fundamental measurements, only ``particles'' flying around in a $3N$-dimensional configuration space, pushed around by a wave function regarded as a real physical field in that space.  In ``spontaneous collapse'' versions, systems are endowed with quantum states that generally evolve unitarily, but from time-to-time collapse without any need for measurement.  In Everettian or ``many-worlds'' quantum mechanics, it is only the world as a whole---they call it a multiverse---that is really endowed with an intrinsic quantum state, and that quantum state evolves deterministically, with only an {\it illusion from the inside\/} of probabilistic ``branching.''

The trouble with all these interpretations as quick fixes for Bell's hard-edged remark is that they look to be just that, {\it really quick fixes}.  They look to be interpretive strategies hardly compelled by the particular details of the quantum formalism, giving only more or less arbitrary appendages to it.  This already explains in part why we have been able to exhibit three such different strategies, but it is worse:  Each of these strategies gives rise to its own set of incredibilities---ones which, if one were endowed with Bell's gift for the pen, one could make look just as silly.  Pilot-wave theories, for instance, give instantaneous action at a distance, but not actions that can be harnessed to send detectable signals.  If so, then what a delicately balanced high-wire act nature presents us with.  Or take the Everettians.  Their world purports to have no observers, but then it has no probabilities either.  What are we then to do with the Born Rule for calculating quantum probabilities?  Throw it away and say it never mattered?  It is true that quite an effort has been made by the Everettians to rederive the rule from {\it decision theory}.  Of those who take the point seriously, some think it works \cite{Wallace09}, some don't \cite{Kent09}.  But outside the {\it sprachspiel\/} who could ever believe?  No amount of sophistry can make ``decision'' anything other than a hollow concept in a predetermined world.


\section{Quantum States Do Not Exist}

There is another lesson from the H1N1 virus.  It is that sometimes immunities can be found in unexpected populations.  To some perplexity, it seems that people over 65---a population usually more susceptible to fatalities with seasonal flu---fare better than younger folk with H1N1.  No one knows exactly why, but the leading theory is that the older population, in its years of other exposures, has developed various latent antibodies.  The antibodies are not perfect, but they are a start.  And so it may be for quantum foundations.

Here, the latent antibody is the concept of {\it information}, and the perfected vaccine, we believe, will arise in part from the theory of single-case, personal probabilities---the branch of probability theory called Bayesianism.  Symbolically, the older population corresponds to some of the very founders of quantum theory (Heisenberg, Pauli, Einstein)\footnote{I feel guilty not mentioning Bohr here, but he so rarely talked directly about quantum states that I fear anything I say would be misrepresentative.\label{FiddleFotts}} and some of the younger disciples of the Copenhagen school (Rudolf Peierls, John Archibald Wheeler, Asher Peres), who, though they disagreed on many details of the vision---{\it Whose information?\ Information about what?}---were unified on one point:  That quantum states are not something out there, in the external world, but instead are expressions of information.  Before there were people using quantum {\it theory\/} as a branch of physics, before they were {\it calculating\/} neutron-capture cross-sections for uranium and working on all the other practical problems the theory suggests, there were no quantum states.  The world may be full of stuff and things of all kinds, but among all the stuff and all the things, there is no unique, observer-independent, {\it quantum-state kind of stuff}.

The immediate payoff of this strategy is that it eliminates the conundrums arising in the various objectified-state interpretations.  A paraphrase of a quote by James Hartle makes the point decisively~\cite{Hartle68}:
\begin{quotation}\small
A quantum-mechanical state being a summary of the observers' information about an individual physical system changes both by
dynamical laws, and whenever the observer acquires new information about the system through the process of measurement.  The existence
of two laws for the evolution of the state vector becomes problematical only if it is believed that the state vector is an objective property of the system.   If, however, the state of a system is defined as a list of [experimental] propositions together with their [probabilities of occurrence], it is not surprising that after a measurement the state must be changed to be in accord with [any] new information.  The ``reduction of the wave packet'' does take place in the consciousness of the observer, not because of any unique physical process which takes place there, but only because the state is a construct of the observer and not an objective property of the physical system.
\end{quotation}
It says that the real substance of Bell's fear is just that, the fear itself.  To succumb to it is to block the way to understanding the theory on its own terms.  Moreover, the shriller notes of Bell's rhetoric are the least of the worries:  The universe didn't have to wait billions of years to collapse its first wave function---wave functions are not part of the observer-independent world.

But this much of the solution is an elderly and somewhat ineffective antibody.  Its presence is mostly a call for more clinical research.  Luckily the days for this are ripe, and it has much to do with the development of the field of quantum information theory in the last 15 years---that is, the multidisciplinary field that has brought about quantum cryptography, quantum teleportation, and will one day bring about full-blown quantum computation.  Terminology can say it all:  A practitioner in this field, whether she has ever thought an ounce about quantum foundations, is just as likely to say ``quantum information'' as ``quantum state'' when talking of any $|\psi\rangle$.  ``What does the quantum teleportation protocol do?''  A now completely standard answer would be: ``It transfers {\it quantum information\/} from Alice's site to Bob's.''  What we have here is a change of mindset~\cite{Fuchs10}.

What the facts and figures, protocols and theorems of quantum information pound home is the idea that quantum states look, act, and feel like information in the technical sense of the word---the sense provided by probability theory and Shannon's information theory.  There is no more beautiful demonstration of this than Robert Spekkens's ``toy model'' for mimicking various features of quantum mechanics \cite{Spekkens07}.  In that model, the ``toys'' are each equipped with four possible mechanical configurations; but the players, the manipulators of the toys, are consistently impeded---for whatever reason!---from having more than one bit of information about each toy's actual configuration. (Or a total of two bits for each two toys, three bits for each three toys, and so on.)  The only things the players can know are their states of uncertainty about the configurations.  The wonderful thing is that these states of uncertainty exhibit many of the characteristics of quantum information: from the no-cloning theorem to analogues of quantum teleportation, quantum key distribution, entanglement monogamy, and even interference in a Mach-Zehnder interferometer.  More than two dozen quantum phenomena are reproduced {\it qualitatively}, and all the while one can always pinpoint the underlying cause of the occurrence:  The phenomena arise in the uncertainties, never in the mechanical configurations.  It is the states of uncertainty that mimic the formal apparatus of quantum theory, not the toys' so-called {\it ontic states\/} (states of reality).

What considerations like this tell the $\psi$-ontologists\footnote{Not to be confused with Scientologists.  This neologism was coined by Chris Granade, a Perimeter Scholars International student at Perimeter Institute, and brought to the author's attention by R.~W. Spekkens, who pounced on it for its beautiful subtlety.}---i.e., those who to attempt to remove the observer too quickly from quantum mechanics by giving quantum states an unfounded ontic status---was well put by Spekkens:
\begin{quote}\small
[A] proponent of the ontic view might argue that the phenomena in question are not mysterious if one abandons certain preconceived notions about physical reality.  The challenge we offer to such a person is to present a few simple physical principles by the light of which all of these phenomena become conceptually intuitive (and not merely mathematical consequences of the formalism) within a framework wherein the quantum state is an ontic state. Our impression is that this challenge cannot be met.  By contrast, a single information-theoretic principle, which imposes a constraint on the amount of knowledge one can have about any system, is sufficient to derive all of these phenomena in the context of a simple toy theory \ldots
\end{quote}
The point is, far from being an appendage cheaply tacked on to the theory, the idea of quantum states as information has a simple unifying power that goes some way toward explaining why the theory has the very mathematical structure it does.\footnote{We say ``goes some way toward'' because, though the toy model makes about as compelling a case as we have ever seen that quantum states are states of information (an extremely valuable step forward), it gravely departs from quantum theory in other aspects. For instance, by its nature, it can give no Bell inequality violations or analogues of the Kochen-Specker noncolorability theorems.  Later sections of this paper will indicate that the cause of the deficit is that the toy model differs crucially from quantum theory in its answer to the question {\it Information about what?}}  By contrast, who could take the many-worlds idea and derive any of the structure of quantum theory out of it?  This would be a bit like trying to regrow a lizard from the tip of its chopped-off tail:  The Everettian conception never purported to be more than a reaction to the formalism in the first place.

There are, however, aspects of Bell's challenge (or at least the mindset behind it), that remain a worry.  And upon these, all could still topple. There are the old questions of {\it Whose information?}\ and {\it Information about what?}---these certainly must be addressed before any vaccination can be declared a success.  It must also be settled whether quantum theory is obligated to give a criterion for what counts as an observer.  Finally, because no one wants to give up on physics, we must tackle head-on the most crucial question of all:  If quantum states are not part of the stuff of the world, then what is?  What sort of stuff does quantum mechanics say the world {\it is\/} made of?

Good immunology does not come easily.  But this much is sure:  The glaringly obvious (that a large part of quantum theory, the central part in fact, is about information) should not be abandoned rashly:  To do so is to lose grip of the theory as it is applied in practice, with no better grasp of reality in return.  If on the other hand, one holds fast to the central point about information, initially frightening though it may be, one may still be able to reconstruct a picture of reality from the unfocused edge of vision.  Often the best stories come from there anyway.

\section{Quantum Bayesianism}

Every area of human endeavor has its bold extremes.  Ones that say, ``If this is going to be done right, we must go this far.  Nothing less will do.''  In probability theory, the bold extreme is the personalist Bayesian account of it \cite{Bernardo94}.  It says that probability theory is of the character of formal logic---a set of criteria for testing consistency.  In the case of formal logic, the consistency is between truth values of propositions.  However logic itself does not have the power to {\it set\/} the truth values it manipulates.  It can only say if various truth values are consistent or inconsistent; the actual values come from another source.  Whenever logic reveals a set of truth values to be inconsistent, one must dip back into the source to find a way to alleviate the discord.  But precisely in which way to alleviate it, logic gives no guidance.  ``Is the truth value for this one isolated proposition correct?''  Logic itself is powerless to say.

The key idea of personalist Bayesian probability theory is that it too is a calculus of consistency (or ``coherence'' as the practitioners call it), but this time for one's decision-making degrees of belief.  Probability theory can only say if various degrees of belief are consistent or inconsistent with each other. The actual beliefs come from another source, and there is nowhere to pin their responsibility but on the agent who holds them.  Dennis Lindley put it nicely in his book {\sl Understanding Uncertainty\/} \cite{Lindley06}:
\begin{quote}\small
The Bayesian, subjectivist, or coherent, para\-digm is egocentric. It is a tale of one person contemplating the world and not wishing to be stupid (technically, incoherent). He realizes that to do this his statements of uncertainty must be probabilistic.
\end{quote}
A probability {\it assignment\/} is a tool an agent uses to make gambles and decisions---it is a tool he uses for navigating life and responding to his environment.  Probability {\it theory\/} as a whole, on the other hand, is not about a single isolated belief, but about a whole mesh of them.  When a belief in the mesh is found to be incoherent with the others, the theory flags the inconsistency.  However, it gives no guidance for how to mend any incoherences it finds.  To alleviate the discord, one can only dip back into the source of the assignments---specifically, the agent who attempted to sum up all his history, experience, and expectations with those assignments in the first place.  This is the reason for the terminology that a probability is a ``degree of belief'' rather than a ``degree of truth'' or ``degree of facticity.''

Where personalist Bayesianism breaks away the most from other developments of probability theory is that it says there are no {\it external\/} criteria for declaring an isolated probability assignment right or wrong.  The only basis for a judgment of adequacy comes from the {\it inside}, from the greater mesh of beliefs the agent may have the time or energy to access when appraising coherence.

It was not an arbitrary choice of words to title the previous section QUANTUM STATES DO NOT EXIST, but a hint of the direction we must take to develop a perfected vaccine.  This is because the phrase has a precursor in a slogan Bruno de Finetti, the founder of personalist Bayesianism, used to vaccinate probability theory itself.  In the preface to his seminal book \cite{DeFinetti90}, de Finetti writes, centered in the page and in all capital letters,
\begin{center}
PROBABILITY$\,$ DOES$\,$ NOT$\,$ EXIST.
\end{center}
It is a powerful statement, constructed to put a finger on the single most-significant cause of conceptual problems in pre-Bayesian probability theory.  A probability is not a solid object, like a rock or a tree that the agent might bump into, but a feeling, an estimate inside himself.

Previous to  Bayesianism, probability was often thought to be a physical property\footnote{Witness Richard von Mises, who even went so far as to write, ``Probability calculus is part of {\it theoretical physics\/} in the same way as classical mechanics or optics, it is an entirely self-contained theory of certain phenomena \ldots''\cite{Mises22}.}---something objective and having nothing to do with decision-making or agents at all.  But when thought so, it could be thought only inconsistently so.  And hell hath no fury like an inconsistency scorned.
The trouble is always the same in all its varied and complicated forms:  If probability is to be a physical property, it had better be a rather ghostly one---one that can be told of in campfire stories, but never quite prodded out of the shadows.  Here's a sample dialogue:
\begin{quote}
\begin{description}
\small
\item[Pre-Bayesian:]  \ Ridiculous, probabilities are without doubt objective.  They can be seen in the relative frequencies they cause.
\item[Bayesian:]  So if $p=0.75$ for some event, after 1000 trials we'll see exactly 750 such events?
\item[Pre-Bayesian:]  \ You might, but most likely you won't see that exactly.  You're just likely to see something close to it.
\item[Bayesian:]  \ Likely?  Close?  How do you define or quantify these things without making reference to your degrees of belief for what will happen?
\item[Pre-Bayesian:]  \ Well, in any case, in the infinite limit the correct frequency will definitely occur.
\item[Bayesian:]  \ How would I know?  Are you saying that in one billion trials I could not possibly see an ``incorrect'' frequency?  In one trillion?
\item[Pre-Bayesian:]  \ OK, you can in principle see an {\it incorrect\/} frequency, but it'd be ever less {\it likely}!
\item[Bayesian:]  \ Tell me once again, what does `likely' mean?
\end{description}
\end{quote}
This is a cartoon of course, but it captures the essence and the futility of every such debate.  It is better to admit at the outset that probability is a degree of belief, and deal with the world on its own terms as it coughs up its objects and events.  What do we gain for our theoretical conceptions by saying that along with each actual event there is a ghostly spirit (its ``objective probability,'' its ``propensity,'' its ``objective chance'') gently nudging it to happen just as it did?  Objects and events are enough by themselves.

Similarly for quantum mechanics.  Here too, if ghostly spirits are imagined behind the actual events produced in quantum measurements, one is left with conceptual troubles to no end.  The defining feature of Quantum Bayesianism~\cite{Caves02,Fuchs02,Fuchs04,Caves07,Fuchs09a,Fuchs10} is that it says along the lines of de Finetti, ``If this is going to be done right, we must go this far.''  Specifically, there can be no such thing as a right and true quantum state, if such is thought of as defined by criteria {\it external\/} to the agent making the assignment:  Quantum states must instead be like personalist, Bayesian probabilities.

The direct connection between the two foundational issues is this.  Quantum states, through the Born Rule, can be used to calculate probabilities.  Conversely, if one assigns probabilities for the outcomes of a well-selected set of measurements, then this is mathematically equivalent to making the quantum-state assignment itself.  The two kinds of assignments determine each other uniquely.  Just think of a spin-$\frac{1}{2}$ system.  If one has elicited one's degrees of belief for the outcomes of a $\sigma_x$ measurement, and similarly one's degrees of belief for the outcomes of $\sigma_y$ and $\sigma_z$ measurements, then this is the same as specifying a quantum state itself:  For if one knows the quantum state's projections onto three independent axes, then that uniquely determines a Bloch vector, and hence a quantum state.  Something similar is true of all quantum systems of all sizes and dimensionality.  There is no mathematical fact embedded in a quantum state $\rho$ that is not embedded in an appropriately chosen set of probabilities.\footnote{See Section \ref{SeekingSICs} where this statement is made precise in all dimensions.}  Thus generally, if probabilities are personal in the Bayesian sense, then so too must be quantum states.

What this buys interpretatively, beside airtight consistency with the best understanding of probability theory, is that it gives each quantum state a home.  Indeed, a home localized in space and time---namely, the physical site of the agent who assigns it!  By this method, one expels once and for all the fear that quantum mechanics leads to ``spooky action at a distance,'' and expels as well any hint of a problem with ``Wigner's friend''~\cite{Wigner71}.  It does this because it removes the very last trace of confusion over whether quantum states might still be objective, agent-independent, physical properties.

The innovation here is that, for most of the history of efforts to take an informational point of view about quantum states, the supporters of the idea have tried to have it both ways:  that on the one hand quantum states are not real physical properties, yet on the other there is a right quantum state independent of the agent after all. For instance, one hears things like, ``The {\it right\/} quantum state is the one the agent should adopt if he had all the information.''  The tension in these two desires leaves their holders open to attack on both flanks and general confusion all around.

Take first instantaneous action at a distance---the horror of this idea is often one of the strongest motivations for those seeking to take an informational stance on quantum states.  But, now an opponent can say:
\begin{quotation}\small
If there is a {\it right quantum state}, then why not be done with all this squabbling and call the state a physical fact to begin with?  It is surely external to the agent if the agent can be wrong about it.  But, once you admit that (and you should admit it), you're sunk: For, now what recourse do you have to declare no action at a distance when a delocalized quantum state changes instantaneously?

\small Here I am with a physical system right in front of me, and though {\it my\/} probabilities for the outcomes of measurements {\it I\/} can do on it might have been adequate a moment ago, there is an objectively better way to gamble {\it now\/} because of something that happened far in the distance?  (Far in the distance and just now.)  How could that not be the signature of action at a distance?  You can try to defend yourself by saying ``quantum mechanics is all about relations''\footnote{A typical example is of a woman traveling far from home when her husband divorces her.  Instantaneously she becomes unmarried---marriage is a relational property, not something localized at each partner.  It seems to be popular to give this example and say, ``Quan\-tum mechanics might be like that.''  The conversation usually stops without elaboration, but let's carry it a little further:  Suppose the woman is right in front of me.  Would the far-off divorce mean that there is instantaneously a different set of probabilities I could use for weighing the consequences of trying to seduce her?  Not at all.  I would have no account to change my probabilities (not for this reason anyway) until I became aware of her changed relation, however long it might take that news to get to me.} or some other feel-good phrase, but I'm talking about measurements right here, in front of me, with outcomes I can see right now.  Ones entering my awareness---not outcomes in the mind of God who can see everything and all relations.  It is that which I am gambling upon with the help of the quantum formalism.  An objectively better quantum state would mean that my gambles and actions, though they would have been adequate a moment ago, are now simply wrong in the eyes of the world---they could have been better.  How could the quantum system in front of me generate outcomes instantiating that declaration without being privy to what the eyes of the world already see?  That's action at a distance, I say, or at least a holism that amounts to the same thing---there's nothing else it could be.
\end{quotation}

Without the protection of truly personal quantum-state assignments, action at a distance is there as doggedly as it ever was.  And things only get worse with ``Wigner's friend'' if one insists there be a {\it right\/} quantum state.  As it turns out, the method of mending this conundrum displays one of the most crucial ingredients of QBism.  Let us put it in plain sight.

``Wigner's friend'' is the story of two agents, Wigner and his friend, and one quantum system---the only deviation we make from a more common presentation\footnote{For instance, \cite{Albert94} is about as common as they get.} is that we put the story in informational terms.  It starts off with the friend and Wigner having a conversation:  Suppose they both agree that some quantum state $|\psi\rangle$ captures their mutual beliefs about the quantum system.\footnote{Being Bayesians, of course, they don't have to agree at this stage---for recall $|\psi\rangle$ is not a physical fact for them, only a catalogue of {\it beliefs}.  But suppose they do agree.} Furthermore suppose they agree that at a specified time the friend will make a measurement on the system of some observable (outcomes $i=1,\ldots,d$).  Finally, they both note that if the friend gets outcome $i$, he will (and should) update his beliefs about the system to some new quantum state $|i\rangle$.  There the conversation ends and the action begins:  Wigner walks away and turns his back to his friend and the supposed measurement.  Time passes to some point beyond when the measurement should have taken place.

\begin{figure}
\begin{center}
\includegraphics[height=2.6in]{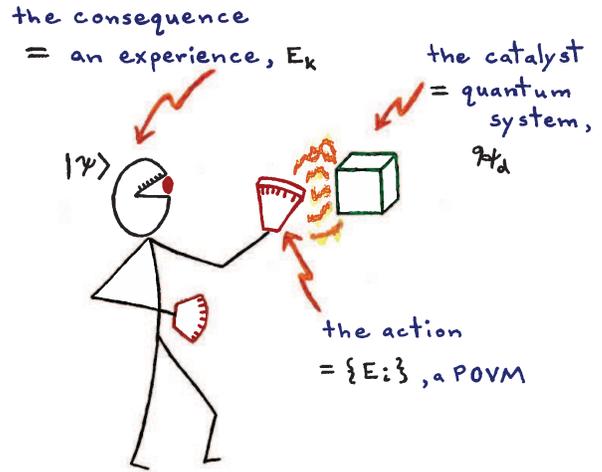}
\bigskip\caption{\small In contemplating a quantum measurement, one makes a conceptual split in the world:  one part is treated as an agent, and the other as a kind of reagent or catalyst (one that brings about change in the agent itself).  The latter is a quantum system of some finite dimension $d$.  A quantum measurement consists first in the agent taking an {\it action\/} on the quantum system.  The action is represented formally by a set of operators $\{E_i\}$---a positive-operator-valued measure. The action generally leads to an incompletely predictable {\it consequence\/} $E_i$ for the agent.  The quantum state $|\psi\rangle$ makes no appearance but in the agent's head; for it captures his degrees of belief concerning the consequences of his actions, and, in contrast to the quantum system itself, has no existence in the external world.  Measurement devices are depicted as prosthetic hands to make it clear that they should be considered an integral part of the agent.  The sparks between the measurement-device hand and the quantum system represent the idea that the consequence of each quantum measurement is a unique creation within the previously existing universe.  Two points are decisive in distinguishing this picture of quantum measurement from a kind of solipsism:  1) The conceptual split of agent and external quantum system: If it were not needed, it would not have been made. 2) Once the agent chooses an action $\{E_i\}$ to take, the particular consequence $E_k$ of it is beyond his control---that is, the actual outcome is not a product of his whim and fancy.}
\end{center}
\end{figure}

What now is the ``correct'' quantum state each agent should have assigned to the quantum system?  We have already concurred that the friend will and
should assign some $|i\rangle$.  But what of Wigner?  If he were to consistently dip into his mesh of beliefs, he would very likely treat his friend as a quantum system like any other:  one with some initial quantum state $\rho$ capturing his (Wigner's) beliefs of {\it it\/} (the friend), along with a linear evolution operator\footnote{We suppose for the sake of introducing less technicality that $U$ is a unitary operation, rather than the more general completely positive trace-preserving linear maps of quantum information theory~\cite{Nielsen00}.  This, however, is not essential to the argument.} $U$ to adjust those beliefs with the flow of time.\footnote{For an explanation of the status of unitary operations from the QBist perspective, as personal judgments directly analogous to quantum states themselves, see Footnote \ref{Quasiclosure} and Refs.~\cite{Fuchs02,Fuchs09a,Leifer06}.}  Suppose this quantum state includes Wigner's beliefs about everything he assesses to be interacting with his friend---in old parlance, suppose Wigner treats his friend as an isolated system.
From this perspective, before any further interaction between himself and the friend or the other system, the quantum state Wigner would assign for the two together would be $U\big(\rho\otimes|\psi\rangle\langle\psi|\big)U^\dagger$ --- most generally an entangled quantum state.  The state of the system itself for Wigner would be gotten from this larger state by a partial trace operation; in any case, it will not be an $|i\rangle$.

Does this make Wigner's new state assignment incorrect?  After all, ``if he had all the information'' (i.e., all the facts of the world) wouldn't that include knowing the friend's measurement outcome? Since the friend should assign some $|i\rangle$, shouldn't Wigner himself (if he had all the information)?  Or is it the friend who is incorrect?  For if the friend had ``all the information,'' wouldn't he say that he is neglecting that Wigner could put the system and himself into the quantum computational equivalent of an iron lung and forcefully reverse the so-called measurement?  I.e., Wigner, if he were sufficiently sophisticated, should be able to force
\begin{equation}
U\big(\rho\otimes|\psi\rangle\langle\psi|\big)U^\dagger\;\;\longrightarrow\;\;\rho\otimes|\psi\rangle\langle\psi|\;.
\label{RatifiedLatified}
\end{equation}
And so the back and forth goes.  Who has the {\it right\/} state of information?  The conundrums simply get too heavy if one tries to hold to an agent-independent notion of correctness for otherwise personalistic quantum states.  The Quantum Bayesian dispels these and similar difficulties of the ``aha, caught you!''~variety by being conscientiously forthright.  {\it Whose information?\/}  ``Mine!''  {\it Information about what?\/}  ``The consequences (for {\it me\/}) of {\it my\/} actions upon the physical system!''  It's all ``I-I-me-me mine,'' as the Beatles sang.

The answer to the first question surely comes as no surprise by now, but why on earth the answer for the second?  ``It's like watching a Quantum Bayesian shoot himself in the foot,'' a friend once said. Why something so egocentric, anthro\-po\-centric, psychology-laden, myopic, and positivistic (we've heard any number of expletives) as {\it the consequences (for me) of my actions upon the system}?  Why not simply say something neutral like ``the outcomes of measurements''?  Or, fall in line with Wolfgang Pauli and say \cite{Pauli94}:
\begin{quote}\small
The objectivity of physics is \ldots\ fully ensured in quantum mechanics in the following sense.  Although in principle, according to the theory, it is in general only the statistics of series of experiments that is determined by laws, the observer is unable, even in the unpredictable single case, to influence the result of his observation---as for example the response of a counter at a particular instant of time.  Further, personal qualities of the observer do not come into the theory in any way---the observation can be made by objective registering apparatus, the results of which are objectively available for anyone's inspection.
\end{quote}
To the uninitiated, our answer for {\it Information about what?}\ surely appears to be a cowardly, unnecessary retreat from realism.  But it is the opposite.  The answer we give is the very injunction that keeps the potentially conflicting statements of Wigner and his friend in check,\footnote{Pauli's statement certainly wouldn't have done that.  Results objectively available for anyone's inspection?  This is the whole issue with ``Wigner's friend'' in the first place.  If both agents could just ``look'' at the counter simultaneously with negligible effect {\it in principle}, we would not be having this discussion.} at the same time as giving each agent a hook to the external world in spite of QBism's egocentric quantum states.

You see, for the QBist, the real world, the one both agents are embedded in---with its objects and events---is taken for granted.  What is not taken for granted is each agent's access to the parts of it he has not touched.  Wigner holds two thoughts in his head: 1) that his friend interacted with a quantum system, eliciting some consequence of the interaction for himself, and 2) after the specified time, for any of Wigner's own further interactions with his friend or system or both, he ought to gamble upon their consequences according to $U\big(\rho\otimes|\psi\rangle\langle\psi|\big)U^\dagger$.  One statement refers to the friend's potential experiences, and one refers to Wigner's own.  So long as it is kept clear that $U\big(\rho\otimes|\psi\rangle\langle\psi|\big)U^\dagger$ refers to the latter---how Wigner should gamble upon the things that might happen to him---making no statement whatsoever about the former, there is no conflict.  The world is filled with all the same things it was before quantum theory came along, like each of our experiences, that rock and that tree, and all the other things under the sun; it is just that quantum theory provides a calculus for gambling on each agent's own experiences---it doesn't give anything else than that.  It certainly doesn't give one agent the ability to conceptually pierce the other agent's personal experience.  It is true that with enough effort Wigner could enact Eq.~(\ref{RatifiedLatified}), causing him to predict that his friend will have amnesia to any future questions on his old measurement results.  But we always knew Wigner could do that---a mallet to the head would have been good enough.

The key point is that quantum theory, from this light, takes nothing away from the usual world of common experience we already know.  It only {\it adds}.\footnote{This point will be much elaborated on in the Section \ref{HSpaceDim}.}  At the very least it gives each agent an extra tool with which to navigate the world.  More than that, the tool is here for a reason.  QBism says when an agent reaches out and touches a quantum system---when he performs a {\it quantum measurement}---that process gives rise to birth in a nearly literal sense.  With the action of the agent upon the system, the no-go theorems of Bell and Kochen-Specker assert that something new comes into the world that wasn't there previously:  It is the ``outcome,'' the unpredictable consequence for the very agent who took the action.  John Archibald Wheeler said it this way, and we follow suit, ``Each elementary quantum phenomenon is an elementary act of `fact creation.'\,''  \cite{Wheeler82c}

With this much, QBism has a story to tell on both quantum {\it states\/} and quantum {\it measurements}, but what of quantum {\it theory\/} as a whole?  The answer is found in taking it as a {\it universal\/} single-user theory in much the same way that Bayesian probability theory is.  It is a users' manual that {\it any\/} agent can pick up and use to help make wiser decisions in this world of inherent uncertainty.\footnote{\label{Macca} Most of the time one sees Bayesian probabilities characterized (even by very prominent Bayesians like Edwin T. Jaynes \cite{Jaynes03}) as measures of ignorance or imperfect knowledge.  But that description carries with it a metaphysical commitment that is not at all necessary for the personalist Bayesian, where probability theory is an extension of logic.  Imperfect knowledge?  It sounds like something that, at least in imagination, could be perfected, making all probabilities zero or one---one uses probabilities only because one does not know the true, pre-existing state of affairs.  Language like this, the reader will notice, is never used in this paper.  All that matters for a personalist Bayesian is that there is {\it uncertainty\/} for whatever reason.  There might be uncertainty because there is ignorance of a true state of affairs, but there might be uncertainty because the world itself does not yet know what it will give---i.e., there is an objective indeterminism.  As will be argued in later sections, QBism finds its happiest spot in an unflinching combination of ``subjective probability'' with ``objective indeterminism.''}  To say it in a more poignant way:  In my case, it is a world in which $I$ am forced to be uncertain about the consequences of most of {\it my\/} actions; and in your case, it is a world in which {\it you\/} are forced to be uncertain about the consequences of most of {\it your\/} actions.  ``And what of God's case?  What is it for him?''  Trying to give {\it him\/} a quantum state was what caused this trouble in the first place!  In a quantum mechanics with the understanding that each instance of its use is strictly single-user---``My measurement outcomes happen right here, to me, and I am talking about my uncertainty of them.''---there is no room for most of the standard, year-after-year quantum mysteries.

\begin{figure}
\begin{center}
\includegraphics[height=2.2in]{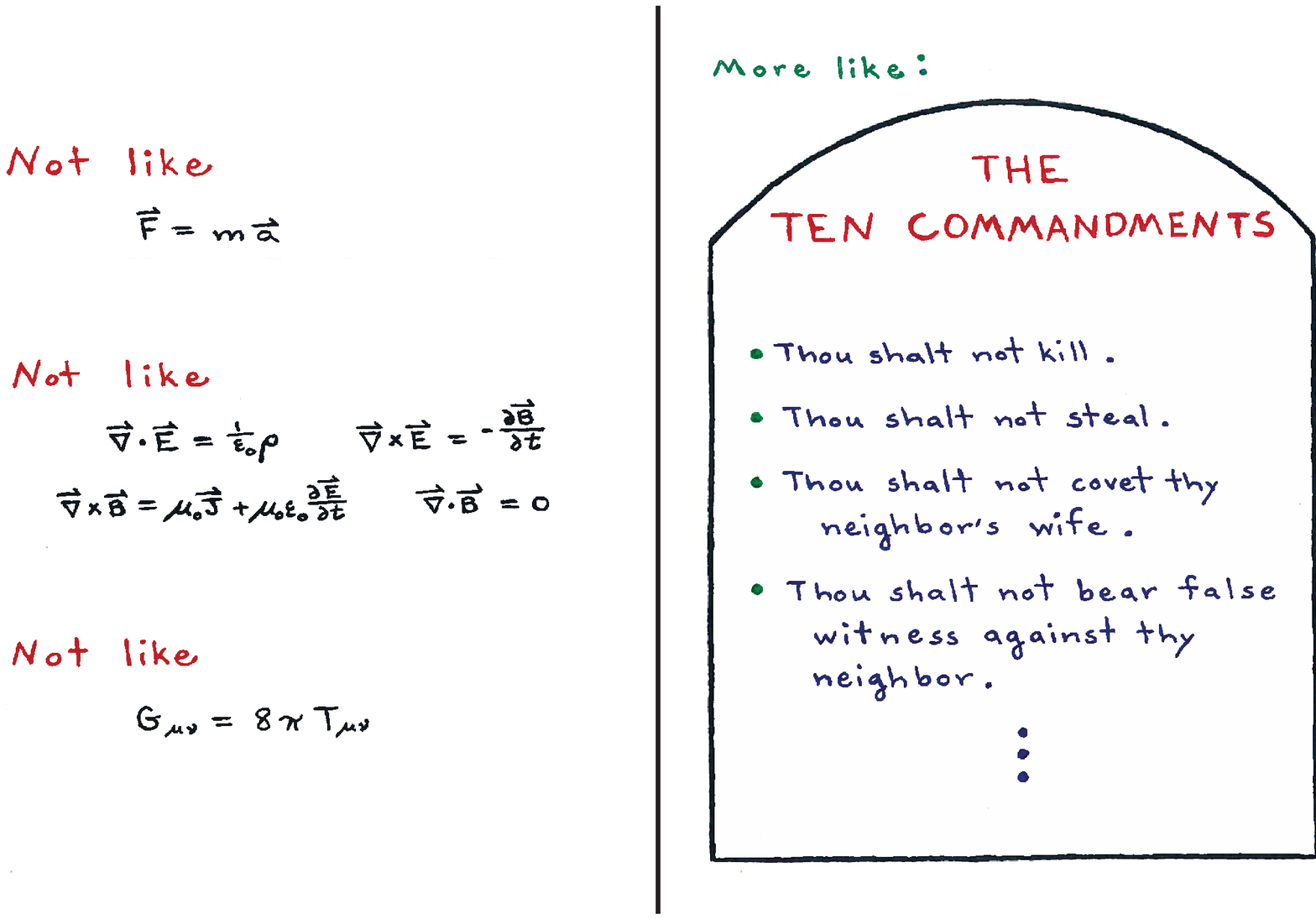}
\caption{\small The Born Rule is not like the other classic laws of physics.  Its normative nature means, if anything, it is more like the Biblical Ten Commandments.  The classic laws on the left give no choice in their statement: If a field is going to be an electromagnetic field at all, it must satisfy Maxwell's equations; it has no choice.  Similarly for the other classic laws.  Their statements are intended to be statements concerning nature {\it just exactly as it is}.  But think of the Ten Commandments.  ``Thou shalt not steal.''  People steal all the time.  The role of the Commandment is to say, ``You have the power to steal if you think you can get away with it, but it's probably not in your best interest to do so.  Something bad is likely to happen as a result.''  Similarly for ``Thou shalt not kill,''\ and all the rest.  It is the worshipper's choice to obey each or not, but if he does not, he ought to count on something potentially bad in return.  The Born Rule guides, ``Gamble in such a way that all your probabilities mesh together through me.''  The agent is free to ignore the advice, but if he does so, he does so at his own peril.}
\end{center}
\end{figure}

The only substantive {\it conceptual\/} issue left before synthesizing a final vaccine\footnote{Not to worry, there are still plenty of technical ones,  as well as plenty more conceptual ones waiting for after the vaccination.} is whether quantum mechanics is obligated to derive the notion of agent for whose aid the theory was built in the first place?  The answer comes from turning the tables:  Thinking of probability theory in the personalist Bayesian way, as an extension of formal logic, would one ever imagine that the notion of an agent, the user of the theory, could be derived out of its conceptual apparatus?  Clearly not.  How could you possibly get flesh and bones out of a calculus for making wise decisions?  The logician and the logic he uses are two different substances---they live in conceptual categories worlds apart.  One is in the stuff of the physical world, and one is somewhere nearer to Plato's heaven of ideal forms.  Look as one might in a probability textbook for the ingredients to reconstruct the reader himself, one will never find them.  So too, the Quantum Bayesian says of quantum theory.

With this we finally pin down the precise way in which quantum theory is ``different in character from any physical theory posed before.''
For the Quantum Bayesian, quantum theory is not something {\it outside\/} probability theory---it is not a picture of the world as it is, as say Einstein's program of a unified field theory hoped to be---but rather it is an {\it addition\/} to probability theory itself.  As probability theory is a {\it normative\/} theory, not saying what one {\it must\/} believe, but offering rules of consistency an agent should strive to satisfy within his overall mesh of beliefs, so it is the case with quantum theory.

To take this substance into one's mindset is all the vac\-ci\-nation one needs against the threat that quantum theory carries something viral for theoretical physics as a whole.  A healthy body is made healthier still.  For with this pro\-tection, we are for the first time in a position to ask, with eyes wide open to what the answer could not be, {\it just what after all is the world made of?}  Far from being the last word on quantum theory, QBism, we believe, is the start of a great adventure.  An adventure full of mystery and danger, with hopes of triumph \ldots\ and all the marks of life.

\section{Seeking SICs -- The Born Rule as Fundamental}

\label{SeekingSICs}

\begin{flushright}
\baselineskip=13pt
\parbox{2.8in}{\baselineskip=13pt\footnotesize

You know how men have always hankered after unlawful magic, and you know
what a great part in magic {\it words\/} have always played. If you
have his name, \ldots\ you can
control the spirit, genie, afrite, or whatever the power may be.
Solomon knew the names of all the spirits, and having their names, he
held them subject to his will.  So the universe has always appeared
to the natural mind as a kind of enigma, of which the key must be
sought in the shape of some illuminating or power-bringing word or
name.  That word names the universe's {\it principle}, and to possess
it is after a fashion to possess the universe itself.\medskip

But if you follow the pragmatic method, you cannot look on any such
word as closing your quest.  You must bring out of each word its
practical cash-value, set it at work within the stream of your
experience.  It appears less as a solution, then, than as a program
for more work, and more particularly as an indication of the ways in
which existing realities may be {\it changed}.\medskip

{\it Theories thus become instruments, not answers to enigmas, in
which we can rest.}  We don't lie back upon them, we move forward,
and, on occasion, make nature over again by their aid.}
\medskip\\
\small --- William James
\end{flushright}

If quantum theory is a user's manual, one cannot forget that the world is its author.  And from its writing style, one may still be able to tell something of the author herself.  The question is how to tease out the psychology of the style, frame it, and identify the underlying motif.

Something that cannot be said of the Quantum Bayesian program is that it has not had to earn its keep in the larger world of quantum interpretations.  Since the beginning, the promoters of the view have been on the run proving technical theorems whenever required to close a gap in its logic or negate an awkwardness induced by its new way of speaking.  It was never enough to ``lie back upon'' the pronouncements:  They had to be shown to have substance, something that would drive physics itself forward.  A case in point is the {\it quantum de Finetti theorem} \cite{Fuchs04,Caves02b}.

This is a theorem that arose from contemplating the meaning of one of the most common phrases of quantum information theory---the unknown quantum state.  The term is ubiquitous:  Unknown quantum states are teleported, protected with quantum error correcting codes, used to check for quantum eavesdropping, and arise in innumerable other applications.  From a Quantum-Bayesian point of view, however, the phrase can only be an oxymoron, something that contradicts itself:  If quantum states are compendia of beliefs, and not states of nature, then the state is known to someone, at the very least the agent who holds it.  But if so, then what are the experimentalists doing when they say they are performing quantum-state tomography in the laboratory?  The very goal of the procedure is to characterize the unknown quantum state a piece of laboratory equipment is repetitively preparing. There is certainly no little agent sitting on the inside of the device devilishly sending out quantum systems representative of his beliefs, and smiling as an experimenter on the outside slowly homes in on those private thoughts through his experiments.  What gives?

The quantum de Finetti theorem is a technical result that allows the story of quantum-state tomography to be told purely in terms of a single agent---namely, the experimentalist in the laboratory.  In a nutshell, the theorem is this.  Suppose the experimentalist walks into the laboratory with the very minimal belief that, of the systems his device is spitting out (no matter how many), he could interchange any two of them and it would not change the statistics he expects for any measurements he might perform.  Then the theorem says ``coherence alone'' requires him to make a quantum-state assignment $\rho^{(n)}$ (for any $n$ of those systems) that can be represented in the form:
\begin{equation}
\rho^{(n)}=\int P(\rho)\, \rho^{\otimes n} d\rho\;,
\label{MushuPork}
\end{equation}
where $P(\rho)\, d\rho$ is some probability measure on the space of single-system density operators and $\rho^{\otimes n}=\rho\otimes\cdots\otimes\rho$ represents an $n$-fold tensor product of identical quantum states. To put it in words, this theorem licenses the experimenter to act {\it as if\/} each individual system has some state $\rho$ unknown to him, with a probability density $P(\rho)$ representing his ignorance of which state is the true one.  But it is only {\it as if}---the only active quantum state in the picture is the one the experimenter (the agent) actually possesses, namely $\rho^{(n)}$.  The right-hand side of Eq.~(\ref{MushuPork}), though necessary among the possibilities, is just one of many representations for $\rho^{(n)}$.  When the experimenter performs tomography, all he is doing is gathering data system-by-system and updating, via Bayes rule \cite{Schack01}, the state $\rho^{(n)}$ to some new state $\rho^{(k)}$ on a smaller number of remaining systems. Particularly, one can prove that this form of quantum-state assignment leads the agent to expect that with more data, he will approach ever more closely a posterior state of the form $\rho^{(k)}=\rho^{\otimes k}$.  This is why one gets into the habit of speaking of tomography as revealing ``the unknown quantum state.''

This example is just one of several \cite{Fuchs04,Caves02c,Appleby05,Leifer06}, and what they all show is that the point of view has some technical crunch\footnote{In fact, the quantum de Finetti theorem has long left its foundational roots behind and found far more widespread recognition with its applications to quantum cryptography \cite{Renner05}.}---it is not just stale, lifeless philosophy.  It stands a chance to ``make nature over again by its aid.''  What better way to master a writer's intentions than to edit her draft and see if she tolerates the changes, admitting in the end that the story flows more easily?

In this regard, no question of QBism tests nature's tolerance more probingly than this.  If quantum theory is so closely allied with probability theory, if it can even be seen as an addition to it, then why is it not written in a language that starts with probability, rather than a language that ends with it?  Why does quantum theory invoke the mathematical apparatus of complex amplitudes, Hilbert spaces, and linear operators?  This brings us to present-day research at Perimeter Institute.

For, actually there are ways to pose quantum theory purely in terms of probabilities---indeed, there are many ways, each with a somewhat different look and feel \cite{Ferrie09}.  The work of W.~K. Wootters is an example, and as he emphasized long ago \cite{Wootters86},
\begin{quotation}\small
It is obviously {\it possible\/} to devise a formulation of quantum mechanics
without probability amplitudes. One is never forced to use any quantities in
one's theory other than the raw results of measurements. However, there is
no reason to expect such a formulation to be anything other than
extremely ugly. After all, probability amplitudes were invented for a reason.
They are not as directly observable as probabilities, but they make the
theory simple. I hope to demonstrate here that one {\it can\/} construct a
reasonably pretty formulation using only probabilities. It may not be quite
as simple as the usual formulation, but it is not much more complicated.
\end{quotation}
What has happened in the intervening years is that the mathematical structures of quantum information theory have grown significantly richer than the ones he had based his considerations on---so much so that we may now be able to optimally re-express the theory.  What was once ``not much more complicated,'' now has the promise of being downright insightful.

The key ingredient is a hypothetical structure called a ``symmetric informationally complete positive-operator-valued measure,'' or SIC (pronounced ``seek'') for short.  This is a set of $d^2$ rank-one projection operators $\Pi_i=|\psi_i\rangle\langle\psi_i|$ on a finite $d$-dimensional Hilbert space such that
\begin{equation}
\big|\langle\psi_i|\psi_j\rangle\big|^2=\frac{1}{d+1}\quad \mbox{whenever} \quad i\ne j\;.
\label{Mojo}
\end{equation}
Because of their extreme symmetry, it turns out that such sets of operators, when they exist, have three very fine-tuned properties: 1) the operators must be linearly independent and span the space of Hermitian operators, 2) there is a sense in which they come as close to an orthonormal basis for operator space as they can (under the constraint that all the elements in a basis be positive semi-definite), and 3) after rescaling, they form a resolution of the identity operator, $I=\sum_i \frac{1}{d}\Pi_i$.

The symmetry, positive semi-definiteness, and properties 1 and 2 are significant because they imply that an arbitrary quantum state $\rho$---pure or mixed---can be expressed as a linear combination of the $\Pi_i$.  Furthermore, the expansion is likely to have some significant features not found in other, more arbitrary expansions.  The most significant of these becomes apparent when one takes property 3 into account.  Because the operators $H_i=\frac1d \Pi_i$ are positive semi-definite and form a resolution of the identity, they can be interpreted as labeling the outcomes of a quantum measurement device---not a standard-textbook, von Neumann measurement device whose outcomes correspond to the eigenvalues of some Hermitian operator, but to a measurement device of the most general variety allowed by quantum theory, the so-called ``positive-operator-valued measures'' (POVMs) \cite{Nielsen00,Peres95}.  Particularly noteworthy is the smooth relation between the probabilities $P(H_i)={\rm tr}\big(\rho H_i\big)$ given by the Born Rule for the outcomes of such a measurement\footnote{There is a slight ambiguity in notation here, as $H_i$ is dually used to denote an operator and an outcome of a measurement. For the sake of simplicity, we hope the reader will forgive this and similar abuses.}
and the expansion coefficients for $\rho$ in terms of the $\Pi_i$:
\begin{equation}
\rho = \sum_{i=1}^{d^2}\left( (d+1)\,P(H_i) - \frac1d \right)\Pi_i\;.
\label{Ralph}
\end{equation}
There are no other operator bases that give rise to such a simple formula connecting probabilities with density operators, and it suggests that this is just the place the Quantum Bayesian should seek his motif.

\begin{figure}
\begin{center}
\includegraphics[width=3.2in]{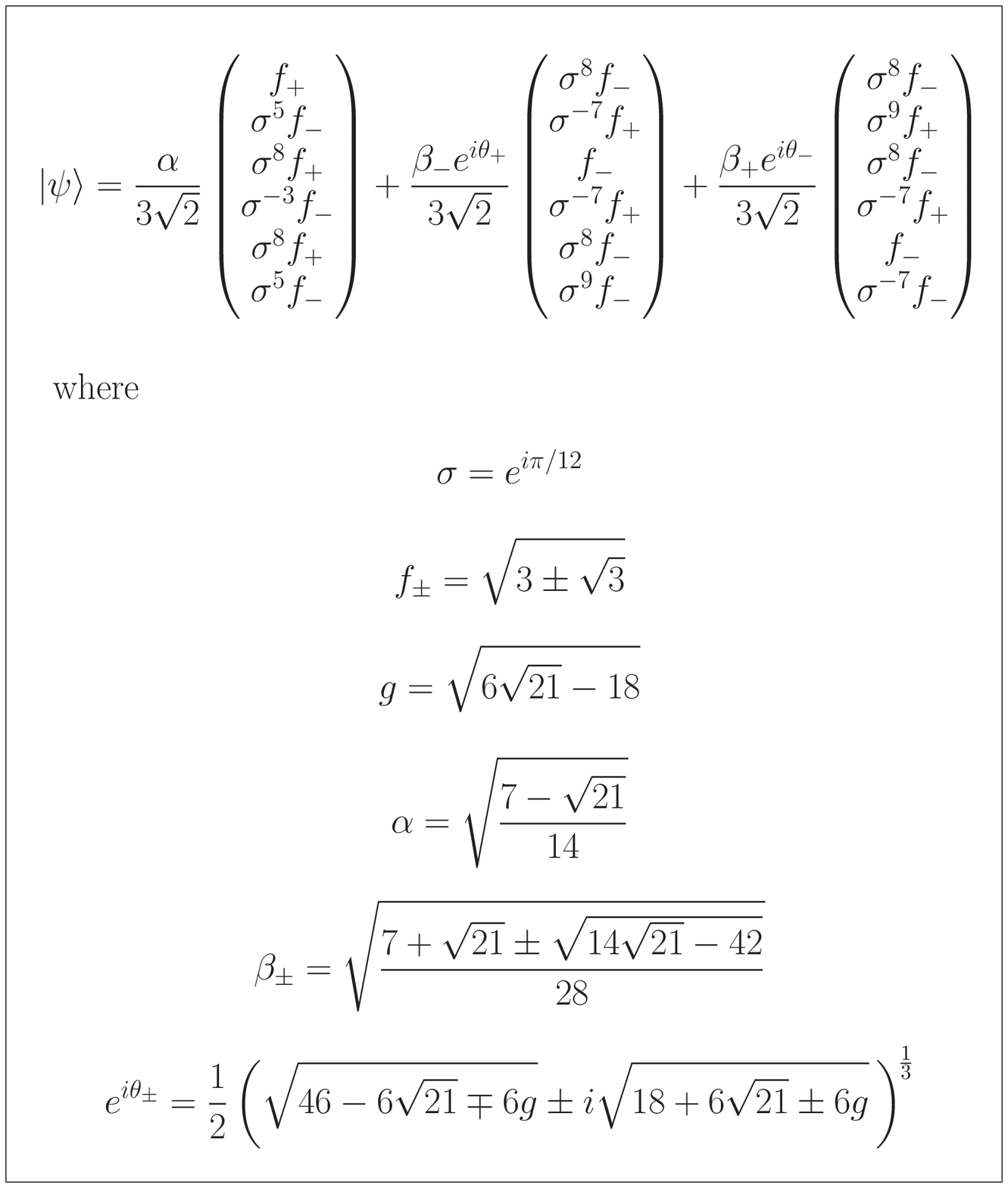}
\bigskip\caption{\small {\bf D.~M. Appleby's ``pencil-and-paper'' SIC in dimension 6}. This is an example of one vector $|\psi\rangle$ among the 36 that go together to form the simplest known SIC in $d=6$.  One of the many problems facing a proof of general SIC existence is that no one has yet latched onto a universal pattern in the existing analytic solutions---every dimension appears to be of a distinct character.}
\end{center}
\end{figure}

Before getting to that, however, we should reveal what is so consternating about the SICs: It is the question of whether they exist at all.  Despite 10 years of growing effort since the definition was first introduced \cite{Zauner99,Caves99,Renes04} (there are now nearly 50 papers on the subject), no one has been able to show that they exist in completely general dimension.  All that is known firmly is that they exist in dimensions 2 through 67 \cite{Scott09}.  Dimensions $2\,$--$\,15$, 19, 24, 35, and 48 are known through direct or computer-automated analytic proof; the remaining solutions are known through numerical simulation, satisfying Eq.~(\ref{Mojo}) to within a precision of $10^{-38}$.  How much evidence is this that SICs exist generally?  The reader must answer this one for himself (certainly there can be no reader-independent answer to something so subjective!), but for the remainder of the article we will proceed as if they do always exist for finite $d$.  At least this is the conceit of our story.  We note in passing, however, that the SIC existence problem is not without wider context:  if they do exist, they solve at least three other (more practical, non-foundational) optimality problems in quantum information theory \cite{Fuchs03,Scott06,Appleby07,Wootters07}---it would be a nasty trick if SICs didn't always exist!

\begin{figure}
\begin{center}
\includegraphics[height=3.3in]{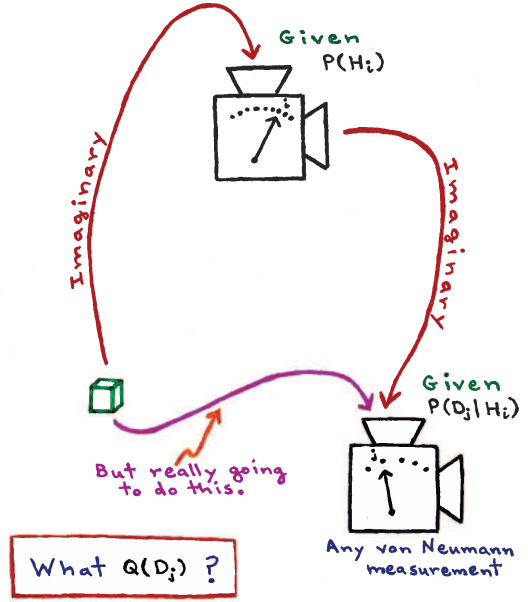}
\bigskip\caption{\small Any quantum measurement can be conceptualized in two ways.
Suppose an arbitrary von Neumann measurement ``on the ground,'' with outcomes $D_j=1,\ldots,d$. Its probabilities $P(D_j)$ can be derived by cascading it with a fixed fiducial SIC measurement ``in the sky'' (of outcomes $H_i=1,\ldots,d^2$). Let $P(H_i)$ and $P(D_j|H_i)$ represent an agent's probabilities, assuming the measurement in the sky is actually performed. The probability $Q(D_j)$ represents instead the agent's probabilities under the assumption that the measurement in the sky is {\it not\/} performed. The Born Rule, in this language, says that $P(D_j)$, $P(H_i)$, and $P(D_j|H_i)$ are related by the Bayesian-style Eq.~(\ref{ScoobyDoo}).}
\end{center}
\end{figure}

So suppose they do.  Thinking of a quantum state as {\it literally\/} an agent's probability assignment for the outcomes of a potential SIC measurement leads to a new way to express the Born Rule for the probabilities associated with any {\it other\/} quantum measurement.  Consider the diagram in Figure 4.  It depicts a SIC measurement ``in the sky,'' with outcomes $H_i$, and any standard von Neumann measurement ``on the ground.''\footnote{Do not, however, let the designation ``SIC sitting in the sky'' make the device seem too exalted and unapproachable.  Actual implementations have already been built for both qubits \cite{Ling06} and qutrits \cite{Medendorp10}.}  For the sake of specificity, let us say the latter has outcomes $D_j=|j\rangle\langle j|$, the vectors $|j\rangle$ representing some orthonormal basis.  We conceive of two possibilities (or two ``paths'') for a given quantum system to get to the measurement on the ground:  ``Path 1'' is that it proceeds directly to the measurement on the ground.  ``Path 2'' is that it proceeds first to the measurement in the sky and only subsequently to the measurement on the ground---the two measurements are cascaded.

Suppose now, we are given the agent's personal probabilities $P(H_i)$ for the outcomes in the sky and his conditional probabilities $P(D_j|H_i)$ for the outcomes on the ground subsequent to the sky.  I.e., we are given the probabilities the agent would assign on the supposition that the quantum system follows Path 2.  Then ``coherence alone'' (in the Bayesian sense) is enough to tell what probabilities $P(D_j)$ the agent should assign for the outcomes of the measurement on the ground---it is given by the Law of Total Probability applied to these numbers:
\begin{equation}
P(D_j)=\sum_i P(H_i) P(D_j|H_i)\;.
\label{Magnus}
\end{equation}
That takes care of Path 2, but what of Path 1?  Is this enough information to recover the probability assignment $Q(D_j)$ the agent would assign for the outcomes on Path 1 via a normal application of the Born Rule?  That is, that
\begin{equation}
Q(D_j)={\rm tr} (\rho D_j)
\label{EmmaPlayStarWars}
\end{equation}
for some quantum state $\rho$?  Maybe, but the answer will clearly not be $P(D_j)$.  One has
\begin{equation}
Q(D_j)\ne P(D_j)\;
\end{equation}
simply because Path 2 is {\it not\/} a coherent process (in the quantum sense!)\ with respect to Path 1---there is a measurement that takes place in Path 2 that does not take place in Path 1.

What is remarkable about the SIC representation is that it implies that, even though $Q(D_j)$ is not equal to $P(D_j)$, it is still a function of it.  Particularly,
\begin{eqnarray}
Q(D_j) &=& (d+1) P(D_j) - 1\nonumber
\\
&=&
(d+1)\sum_{i=1}^{d^2} P(H_i) P(D_j|H_i) - 1\;.
\label{ScoobyDoo}
\end{eqnarray}
The Born Rule is nothing but a kind of Quantum Law of Total Probability!  No complex amplitudes, no operators---only probabilities in, and probabilities out.  Indeed, it is seemingly just a rescaling of the old law, Eq.~(\ref{Magnus}).  And in a way it is.

{\it But beware}:  One should not interpret Eq.~(\ref{ScoobyDoo}) as invalidating probability theory itself in any way:  For the old Law of Total Probability has no jurisdiction in the setting of our diagram, which compares a ``factual'' experiment (Path 1) to a ``counterfactual'' one (Path 2).\footnote{Indeed, as we have emphasized, there is a trace of a very old antibody in QBism.  While writing this essay, it came to light in the nice historical study of Ref.~\cite{Bacciagaluppi09} that Born and Heisenberg, {\it already at the 1927 Solvay conference}, refer to the calculation $|c_n(t)|^2=\left|\sum_m S_{mn}(t)c_m(0)\right|^2$ and say, ``it should be noted that this `interference' does not represent a contradiction with the rules of the probability calculus, that is, with the assumption that the $|S_{nk}|^2$ are quite usual probabilities.'' Their reasons for saying this may have been different from our own, but at least they had come this far.}  Indeed as any Bayesian would emphasize, if there is a distinguishing mark in one's considerations---say, the fact of two distinct experiments, not one---then one ought to take that into account in one's probability assignments (at least initially so).  Thus there is a hidden, or at least suppressed, condition in our notation:  Really we should have been writing the more cumbersome, but honest, expressions $P(H_i|{\mathcal E}_2)$, $P(D_j|H_i,{\mathcal E}_2)$, $P(D_j|{\mathcal E}_2)$, and $Q(D_j|{\mathcal E}_1)$ all along.  With this explicit, it is no surprise that,
\begin{equation}
Q(D_j|{\mathcal E}_1)\ne\sum_iP(H_i|{\mathcal E}_2)P(D_j|H_i,{\mathcal E}_2)\;.
\end{equation}
The message is that quantum theory supplies some\-thing---a new form of ``Bayesian coherence,'' though empirically based (as quantum theory itself is)---that raw probability theory does not.  The Born Rule in these lights is an addition to Bayesian probability, not in the sense of a supplier of some kind of more-objective probabilities, but in the sense of giving extra normative rules to guide the agent's behavior when he interacts with the physical world.

It is a normative rule for reasoning about the consequences of one's proposed actions in terms of the potential consequences of an explicitly counterfactual action.  It is like nothing else physical theory has contemplated before.  Seemingly at the heart of quantum mechanics from the QBist view is a statement about the impact of {\it counterfactuality\/}.  The impact parameter is metered by a single, significant number associated with each physical system---its Hilbert-space dimension $d$.  The larger the $d$ associated with a system, the more $Q(D_j)$ must deviate from $P(D_j)$.  Of course this point must have been implicit in the usual form of the Born Rule, Eq.~(\ref{EmmaPlayStarWars}).  What is important from the QBist perspective, however, is how the new form puts the significant parameter front and center, displaying it in a way that one ought to nearly trip over.

Understanding this as the goal helps pinpoint the role of SICs in our considerations.  The issue is not that quantum mechanics {\it must\/} be rewritten in terms of SICs, but that it {\it can\/} be.\footnote{If everything goes right, that is, and the damned things actually exist in all dimensions!}  Certainly no one is going to drop the usual operator formalism and all the standard methods learned in graduate school to do their workaday calculations in SIC language exclusively.  It is only that the SICs form an ideal coordinate system for a particular problem (an important one to be sure, but nonetheless a particular one)---the problem of {\it interpreting\/} quantum mechanics.  The point of all the various representations of quantum mechanics (like the various quasi-probability representations of \cite{Ferrie09}, the Heisenberg and Schr\"odinger pictures, and even the path-integral formulation) is that they give a means for isolating one or another aspect of the theory that might be called for by a problem at hand.  Sometimes it is really important to do so, even for deep conceptual issues and even if all the representations are logically equivalent.\footnote{Just think of the story of Eddington-Finkelstein coordinates in general relativity.  Once upon a time it was not known whether a Schwarzschild black hole might have, beside its central singularity, a singularity in the gravitational field at the event horizon.  Apparently it was a heated debate, yes or no. The issue was put to rest, however, with the development of the coordinate system.  It allowed one to write down a solution to the Einstein equations in a neighborhood of the horizon and check that everything was alright after all.}  In our case, we want to bring into plain view the idea that quantum mechanics is an {\it addition\/} to Bayesian probability theory---not a generalization of it \cite{Bub07}, not something orthogonal to it altogether \cite{Jozsa04}, but an addition.  With this goal in mind, the SIC representation is a particularly powerful tool.  Through it, one sees the Born Rule as a {\it functional\/} of a usage of the Law of Total Probability that one would have made in another (counterfactual) context.\footnote{\label{Quasiclosure} Furthermore it is similarly so of unitary time evolution in a SIC picture.  To explain what this means, let us change considerations slightly and make the measurement on the ground a unitarily rotated version of the SIC in the sky.  This contrasts with the von Neumann measurement we have previously restricted the ground measurement to be.  In this setting, $D_j=\frac{1}{d}U\Pi_j U^\dagger$, which in turn implies a slight modification to Eq.~(\ref{ScoobyDoo}), \begin{equation} Q(D_j) = (d+1)\sum_{i=1}^{d^2} P(H_i) P(D_j|H_i) - \frac{1}{d}\;, \label{Zizek}\end{equation} for the probabilities on the ground.  Note what this is saying!  As the Born Rule is a functional of the Law of Total Probability, unitary time evolution is a functional of it as well.  For, if we thought in terms of the Sch\"odinger picture, $P(H_i)$ and $Q(D_j)$ would be the SIC representations for the initial and final quantum states under an evolution given by $U^\dagger$. The similarity is no accident.  This is because in both cases the conditional probabilities $P(D_j|H_i)$ completely encode the identity of a measurement on the ground.%

Moreover, it makes abundantly clear another point of QBism that has not been addressed so much in the present paper.  Since a personalist Bayesian cannot turn his back on the clarification that {\it all\/} probabilities are personal judgments, placeholders in a calculus of consistency, he certainly cannot turn his back on the greater lesson Eqs.~(\ref{ScoobyDoo}) and (\ref{Zizek}) are trying to scream out.   Just as quantum states $\rho$ are personal judgments $P(H_i)$, quantum measurement operators $D_j$ and unitary time evolutions $U$ are personal judgments too---in this case $P(D_j|H_i)$.  The only distinction is the technical one, that one expression is an unconditioned probability, while the other is a collection of conditionals.  Most importantly, it settles the age-old issue of why there should be two kinds of state evolution at all.  When Hartle wrote, ``A quantum-mechanical state being a summary of the observers' information about an individual physical system changes both by dynamical laws, and whenever the observer acquires new information about the system through the process of measurement,'' what is his dynamical law making reference to?  There are not two things that a quantum state can do, only one:  Strive to be consistent with all the agent's other probabilistic judgments on the consequences of his actions, factual and counterfactual.}  The SICs emphasize and make this point clear.  At the end of the day however, after all the foundational worries of quantum theory are finally overcome, the SICs might in principle be thrown away, just as the scaffolding surrounding any finished construction would be.

Much of the most intense research of Perimeter Institute's QBism group is currently devoted to seeing how much of the essence of quantum theory is captured by Eq.~(\ref{ScoobyDoo}).  For instance, one way to approach this is to take Eq.~(\ref{ScoobyDoo}) as a fundamental axiom and ask what further assumptions are required to recover all of quantum theory?  To give some hint of how a reconstruction of quantum theory might proceed along these lines, note Eq.~(\ref{Ralph}) again.  What it expresses is that any quantum state $\rho$ can be reconstructed from the probabilities $P(H_i)$ the state $\rho$ gives rise to.  This, however, does not imply that plugging just any probability distribution $P(H_i)$ into the equation will give rise to a valid quantum state.  A general probability distribution $P(H_i)$ in the formula will lead to a Hermitian operator of trace one, but it may not lead to an operator with nonnegative eigenvalues.  Indeed it takes further restrictions on the $P(H_i)$ to make this true.  That being the case, the Quantum Bayesian starts to wonder if these restrictions might arise from the requirement that Eq.~(\ref{ScoobyDoo}) simply always make sense.  For note, if $P(D_j)$ is too small, $Q(D_j)$ will go negative; and if $P(D_j)$ is too large, $Q(D_j)$ will become larger than 1.  So, $P(D_j)$ must be restricted.  But that in turn forces the set of valid $P(H_i)$ to be restricted as well.  And so the argument goes.  For sure, some amount of quantum theory (and maybe all of it) is reconstructed in this fashion \cite{Fuchs09a,Appleby09a,Fuchs09b,Appleby09b}.

Another exciting development comes from loosening the form of Eq.~(\ref{ScoobyDoo}) to something more generic:
\begin{equation}
Q(D_j)=\alpha\sum_{i=1}^n P(H_i) P(D_j|H_i) - \beta\;,
\end{equation}
where there is {\it initially\/} no assumed relation between $\alpha$, $\beta$, and $n$ as there is in Eq.~(\ref{ScoobyDoo}).  Then, under a few further conditions with only the faintest hint of quantum theory in them---for instance, that there should exist measurements on the ground for which, under appropriate conditions, one can have certainty for their outcomes---one immediately gets a significantly more restricted form for this relation:
\begin{equation}
Q(D_j)=\left(\frac{1}{2} qd+1\right)\sum_{i=1}^{n} P(H_i) P(D_j|H_i) - \frac12 q\;,
\label{ExNihiloOmnia}
\end{equation}
where very interestingly the parameters $q$ and $d$ can only take on integer values, $q=0,1,2,\ldots,\infty$ and $d=2,3,4,\ldots,\infty$, and $n=\frac12 qd(d-1) + d$.

The $q=2$ case can be identified with the quantum me\-chanical one we have seen before.  On the other hand, the $q=0$ case can be identified with the usual vision of the classical world:  A world where counterfactuals simply do not matter, for the world just ``is.''  In this case, an agent is well advised to take $Q(D_j)=P(D_j)$, meaning that there is no operational distinction between experiments ${\mathcal E}_1$ and ${\mathcal E}_2$ for him. It should not be forgotten however, that this rule, trivial though it looks, is still an addition to raw probability theory.  It is just one that meshes well with what had come to be expected by most classical physicists.  To put it yet another way, in the $q=0$ case, the agent says to himself that the fine details of his actions do not matter. This to some extent authorizes the view that observation is a passive process in principle---again the classical worldview.  Finally, the cases $q=1$ and $q=4$, though not classical, track still other structures that have been explored previously:  They correspond to what the Born Rule would look like if alternate versions of quantum mechanics, those over real~\cite{Stueckelberg60} and quaternionic~\cite{AdlerBook} vector spaces, were expressed in the equivalent of SIC terms.\footnote{The equivalent of SICs (i.e., informationally complete sets of equiangular projection operators) certainly do {\it not\/} exist in general dimensions for the real-vector-space case---instead these structures only exist in a sparse set of dimensions, $d=2,3,7,23,\ldots\,$. With respect to the quaternionic theory, it appears from numerical work that they do not generally exist in that setting either \cite{Khatirinejad08}.  Complex quantum mechanics, like baby bear's possessions, appears to be just right.}

Formula (\ref{ExNihiloOmnia}) from the general setting indicates more strongly than ever that it is the role of {\it dimension\/} that is key to distilling the motif of our user's manual.  Quantum theory, seen as a normative addition to probability theory, is just one theory (the second rung above classical) along an infinite hierarchy.  What distinguishes the levels of this hierarchy is the strength $q$ with which dimension ``couples'' the two paths in our diagram of Figure 2.  It is the strength with which we are compelled to deviate from the Law of Total Probability when we transform our thoughts from the consequences of counterfactual actions upon a $d$'s worth of the world's stuff to the consequences of our factual ones.  Settling upon $q=2$ (i.e., settling upon quantum theory itself) sets the strength of the coupling, but the $d$ variable remains.  Different systems, different $d$, different deviations from a naive application of the Law of Total Probability.

In some way yet to be fully fleshed out, each quantum system seems to be a seat of active coun\-ter\-factuality and possibility, whose outward effect is as an ``agent of change'' for the parts of the world that come into contact with it.  Observer and system, ``agent and reagent,'' might be a way to put it.  Perhaps no metaphor is more pregnant for QBism's next move than this:  If a quantum system is comparable to a chemical reagent, then $d$ is comparable to a valence.  But valence for {\it what\/} more exactly?

\section{The Essence of Bell's Theorem, QBism Style}

\label{BellTheorem}

It is easy enough to {\it say\/} that a quantum system (and hence each piece of the world) is a ``seat of possibility.''  In a spotty way, certain philosophers have been saying similar things for 150 years.  What is unique about quantum theory in the history of thought is the way in which its mathematical structure has pushed this upon us to our very surprise.  It wasn't that all these grand statements on the philosophical structure of the world were built into the formalism, but that the formalism reached out and shook its users until they opened their eyes.  Bell's theorem and all its descendants are examples of that.

So when the users opened their eyes, what did they see?  From the look of several recent prominent expositions on the subject \cite{Gisin09,Albert09,Norsen06}, it was ``nonlocality everlasting!''  That the world really is full of spooky action at a distance---live with it and love it.  But conclusions drawn from even the most rigorous of theorems can only be {\it additions\/} to one's prior understanding and beliefs when the theorems do not contradict those beliefs flat out.  Such was the case with Bell's theorem.  It has just enough room in it to not contradict a misshapen notion of probability, and that is the hook and crook that the lovers of Star Trek have thrived on.  The Quantum Bayesian, however, with a different understanding of probability and a commitment to the idea that quantum measurement outcomes are personal, draws quite a different conclusion from the theorem.  In fact it is a conclusion from the far opposite end of the spectrum:  It tells of a world unknown to most monist and rationalist philosophies:  The universe, far from being one big nonlocal block, should be thought of as a thriving community of marriageable, but otherwise autonomous entities.   That the world should violate Bell's theorem remains, even for QBism, the deepest statement ever drawn from quantum theory.  It says that quantum measurements are {\it moments of creation}.

This language has already been integral to our presentation, but seeing it come about in a formalism-driven way like Bell's makes the issue particularly vivid.  Here we devote some effort to showing that the language of creation is a consequence of three things:  1) the quantum formalism, 2) a personalist Bayesian interpretation of probability, and 3) the elementary notion of what it means to be two objects rather than one.  We do not do it however with Bell's theorem precisely, but with an argument that more directly implicates the EPR ``criterion of reality'' as the source of trouble with quantum theory.  The thrust of it is that it is the EPR criterion that should be jettisoned, not locality.

Our starting point is like our previous setup---an agent and a system---but this time we make it two systems:  One of them, the left-hand one, is ready. The other, the right-hand one, is waiting.  The agent will eventually measure each in turn.\footnote{It should be noted how we depart from the usual presentation here:  There is only the single agent and his two systems.  There is no Alice and Bob accompanying the two systems.}  Simple enough to say, but things get hung at the start with the issue of what is meant by ``two systems?''  A passage from a 1948 paper of Einstein \cite{Einstein48} captures the essential issue well:
\begin{quotation}\small
If one asks what is characteristic of the realm of physical ideas independently of the quantum-theory, then above all the following attracts our attention: the concepts of physics refer to a real external world, i.e., ideas are posited of things that claim a ``real existence'' independent of the perceiving subject (bodies, fields, etc.), and these ideas are, on the one hand, brought into as secure a relationship as possible with sense impressions. Moreover, it is characteristic of these physical things that they are conceived of as being arranged in a space-time continuum. Further, it appears to be essential for this arrangement of the things introduced in physics that, at a specific time, these things claim an existence independent of one another, insofar as these things ``lie in different parts of space.'' Without such an assumption of the mutually independent existence (the ``being-thus'') of spatially distant things, an assumption which originates in everyday thought, physical thought in the sense familiar
to us would not be possible. Nor does one see how physical laws could be formulated and tested without such a clean separation. \ldots 

For the relative independence of spatially distant things (A and B), this idea is characteristic: an external influence on A has no {\it immediate\/} effect on B; this is known as the ``principle of local action,'' \ldots. 
The complete suspension of this basic principle would make impossible the idea of (quasi-) closed systems and, thereby, the establishment of empirically testable laws in the sense familiar to us.
\end{quotation}
We hope it is clear to the reader by now that QBism concurs with every bit of this.  Quantum states may not be the stuff of the world, but QBists never shudder from positing quantum systems as ``real existences'' external to the agent.  And just as the agent has learned from long, hard experience that he cannot reach out and touch anything but his immediate surroundings, so he imagines of every quantum system, one to the other.  What is it that A and B are spatially distant things but that they are causally independent?

This notion, in Einstein's hands,\footnote{Beware!  This is {\it not\/} to say in the hands of EPR---Einstein, Podolsky, and Rosen.  The present argument is not their argument.  For a discussion of Einstein's dissatisfaction with the one appearing in the EPR paper itself, see \cite{Fine96}.} led to one of the nicest, most direct arguments that quantum states cannot be states of reality, but must be something {\it more like\/} states of information, knowledge, expectation, or belief \cite{Harrigan07}.  The argument is important---let us repeat the whole thing from Einstein's most thorough version of it \cite{Einstein49}.  It more than anything sets the stage for a QBist development of a Bell-style contradiction.
\begin{quotation}\small
Physics is an attempt conceptually to grasp reality as it is thought independently of its being observed.  In this sense on speaks of ``physical reality.'' In pre-quantum physics there was no doubt as to how this was to be understood.  In Newton's theory reality was determined by a material point in space and time; in Maxwell's theory, by the field in space and time.  In quantum mechanics it is not so easily seen.  If one asks: does a $\psi$-function of the quantum theory represent a real factual situation in the same sense in which this is the case of a material system of points or of an electromagnetic field, one hesitates to reply with a simple ``yes'' or ``no''; why?  What the $\psi$-function (at a definite) time asserts, is this:  What is the probability for finding a definite physical magnitude $q$ (or $p$) in a definitely given interval, if I measure it at time $t$?  The probability is here to be viewed as an empirically determinable, therefore certainly as a ``real'' quantity which I may determine if I create the same $\psi$-function very often and perform a $q$-measurement each time.  But what about the single measured value of $q$?  Did the respective individual sys\-tem have this $q$-value even before this measurement?  To this question there is no definite answer within the framework of the theory, since the measurement is a pro\-cess which implies a finite disturbance of the sys\-tem from the outside; it would therefore be thinkable that the system obtains a definite numerical value for $q$ (or $p$) the measured numerical value, only through the measurement itself.  For the further discussion I shall assume two physicists $A$ and $B$, who represent a different conception with reference to the real situation as described by the $\psi$-function.
\begin{itemize}
\item[$A$.] \small The individual system (before the measurement) has a definite value of $q$ (or $p$) for all variables of the system, and more specifically, {\it that\/} value which is determined by a measurement of this var\-i\-able.  Proceeding from this conception, he will state: The $\psi$-function is no exhaustive description of the real situation of the system but an incomplete description; it expresses only what we know on the basis of former measurements concerning the system.
\item[$B$.] The individual system (before the measurement) has no definite value of $q$ (or $p$).  The value of the measurement only arises in cooperation with the unique probability which is given to it in view of the $\psi$-function only through the act of measurement itself.  Proceeding from this conception, he will (or, at least, he may) state:  The $\psi$-function is an exhaustive description of the real situation of the system.
\end{itemize}
We now present to these two physicists the following instance:  There is to be a system which at the time $t$ of our observation consists of two partial systems $S_1$ and $S_2$, which at this time are spatially separated and (in the sense of classical physics) are without significant reciprocity.  The total system is to be completely described through a known $\psi$-function $\psi_{12}$ in the sense of quantum mechanics.  All quantum theoreticians now agree upon the following:  If I make a complete measurement of $S_1$, I get from the results of the measurement and from $\psi_{12}$ an entirely definite $\psi$-function $\psi_2$ of the system $S_2$.  The character of $\psi_2$ then depends upon {\it what kind\/} of measurement I undertake on $S_1$.

Now it appears to me that one may speak of the real factual situation of the partial system $S_2$. Of this real factual situation, we know to begin with, before the measurement of $S_1$, even less than we know of a system described by the $\psi$-function.
But on one supposition we should, in my opinion, absolutely hold fast: The real factual situation of the system $S_2$ is independent of what is done with the system $S_1$, which is spatially separated from the former.  According to the type of measurement which I make of $S_1$, I get, however, a very different $\psi_2$ for the second partial system. Now, however, the real situation of $S_2$ must be independent of what happens to $S_1$.  For the same real situation of $S_2$ it is possible therefore to find, according to one's choice, different types of $\psi$-function. \ldots

If now the physicists, $A$ and $B$, accept this consideration as valid, then $B$ will have to give up his position that the $\psi$-function constitutes a complete description of a real factual situation.  For in this case it would be impossible that two different types of $\psi$-functions could be coordinated with the identical factual situation of $S_2$.
\end{quotation}
Aside from asserting a frequentistic conception of probability, the argument is nearly perfect.\footnote{You see, there really was a reason for including Einstein with Heisenberg, Pauli, Peierls, Wheeler, and Peres at the beginning of the article.  Still, please reread Footnote \ref{Macca}.}  It tells us one important reason why we should not be thinking of quantum states as the $\psi$-ontologists do.  Particularly, it is one we should continue to bear in mind as we move to a Bell-type setting:  Even there, there is no reason to waiver on its validity.  It may be true that Einstein implicitly equated ``incomplete description'' with ``there must exist a hidden-variable account'' (though we do not think he did), but the argument as stated neither stands nor falls on this issue.

There is, however, one thing that Einstein does miss in his argument, and this is where the structure of Bell's thinking steps in.  Einstein says, ``to this question there is no definite answer within the framework of the theory'' when speaking of whether quantum measurements are ``generative'' or simply ``revealing'' of their outcomes.  If we accept everything he has already said, then with a little clever combinatorics and geometry one can indeed settle the question.

Let us suppose that the two spatially separated systems in front of the agent are two ququarts (i.e., each system is associated with a four-dimensional Hilbert space ${\mathcal H}_4$), and that the agent ascribes a maximally entangled state to the pair, i.e., a state $|\psi\rangle$ in ${\mathcal H}_4\otimes{\mathcal H}_4$ of the form,
\begin{equation}
|\psi\rangle=\frac{1}{2}\sum_{i=1}^4|i\rangle|i\rangle\;.
\end{equation}
Then we know that there exist pairs of measurements, one for each of the separate systems, such that if the outcome of one is known (whatever the outcome), one will thereafter make a probability-one statement concerning the outcome of the other.  For instance, if a nondegenerate Hermitian operator $H$ is measured on the left-hand system, then one will thereafter ascribe a probability-one assignment for the appropriate outcome of the transposed operator $H^{\rm\small T}$ on the right-hand system.  What this means for a Bayesian agent is that after performing the first measurement he will bet his life on the outcome of the second.

But how could that be if he has already recognized two systems with no instantaneous causal influence between each other?  Mustn't it be that the outcome on the right-hand side is ``already there'' simply awaiting confirmation or registration?  It would seem Einstein's physicist $B$ is already living in a state of contradiction.

Indeed it must be this kind of thinking that led Einstein's collaborators Podolsky and Rosen to their famous sufficient criterion for an ``element of [preexistent] reality'' \cite{Fine96}:
\begin{quote}\small
If, without in any way disturbing a system, we can predict with certainty (i.e., with probability equal to unity) the value of a physical quantity, then there exists an element of reality corresponding to that quantity.
\end{quote}
Without doubt, no personalist Bayesian would ever utter such a notion:  Just because he believes something with all his heart and soul and would gamble his life on it, it would not make it necessarily so by the powers of nature---even a probability-one assignment is a state of belief for the personalist Bayesian.  But he might still entertain something not unrelated to the EPR criterion of reality.  Namely, that believing a particular outcome will be found with certainty on a causally disconnected system entails that one {\it also\/} believes the outcome to be ``already there'' simply awaiting confirmation.

But it is not so, and the Quantum Bayesian has already built this into his story of measurement.  Let us show this presently\footnote{Overall this particular technique has its roots in Stairs \cite{Stairs83}, and seems to bear some resemblance to the gist of Conway and Kochen's ``Free Will Theorem'' \cite{Conway06,Conway09}. } by combining all the above with a beautifully simple Kochen-Specker style construction discovered by Cabello, Estebaranz, and Garc\'{\i}a-Alcaine (CEGA) \cite{Cabello97}.  Imagine some measurement $H$ on the left-hand system; we will denote its potential outcomes as a column of letters, like this
\begin{equation}
\veec abcd
\end{equation}
Further, since there is a fixed transformation taking any $H$ on the left-hand system to a corresponding $H^{\rm\small T}$ on the right-hand one, there is no harm in identifying the notation for the outcomes of both measurements.  That is to say, if the agent gets outcome $b$ (to the exclusion of $a$, $c$, and $d$) for $H$ on the left-hand side, he will make a probability-one prediction for $b$ on the right-hand side, even though that measurement strictly speaking is a different one, namely $H^{\rm\small T}$.  If the agent further subscribes to (our Bayesian variant of) the EPR criterion of reality, he will say that he believes $b$ to be TRUE of the right-hand system as an element of reality.

Now let us consider two possible measurements, $H_1$ and $H_2$ for the left-hand side, with potential outcomes
\begin{equation}
\veec abcd \quad\qquad\mbox{and}\qquad\quad \veec efgh
\end{equation}
respectively.  Both measurements cannot be performed at once, but it might be the case that if the agent gets a specific outcome for $H_1$, say $c$ particularly, then not only will he make a probability-one assignment for $c$ in a measurement of $H_1^{\rm\small T}$ on the right-hand side, but also for $e$ in a measurement of $H_2^{\rm\small T}$ on it.  Similarly, if $H_2$ were measured on the left, getting an outcome $e$; then he will make a probability-one prediction for $c$ in a measurement of $H_1^{\rm\small T}$ on the right.  This would come about if $H_1$ and $H_2$ (and consequently $H_1^{\rm\small T}$ and $H_2^{\rm\small T}$) share a common eigenvector.  Supposing so and that $c$ was actually the outcome for $H_1$ on the left, what conclusion would the EPR criterion of reality draw?  It is that both $c$ and $e$ are elements of reality on the right, and none of $a$, $b$, $d$, $f$, $g$, or $h$ are.  Particularly, since the right-hand side could not have known whether $H_1$ or $H_2$ was measured on the left, whatever $c$ and $e$ stands for, it must be the same thing, the same property.  In such a case, we discard the extraneous distinction between $c$ and $e$ in our notation and write
\begin{equation}
\veec abcd \quad\qquad\mbox{and}\qquad\quad \veec cfgh
\end{equation}
for the two potential outcome sets for a measurement on the right.

We now have all the notational apparatus we need to have some fun.  The genius of CEGA was that they were able to find a set of nine ``interlocking'' Hermitian operators $H_1$, $H_2$, \ldots, $H_9$ for the left, whose set of potential outcomes for the corresponding operators on the right would look like this:
\begin{equation}
\veec abcd \qquad \veec aefg \qquad \veec hicj\qquad \veec hkgl\qquad \veec bemn\qquad \veec ikno\qquad \veec pqdj\qquad \veec prfl\qquad \veec qrmo \label{BurlIves}
\end{equation}
Take the second column as an example.  It means that if $H_2$ were measured on the left-hand system, only one of $a$, $e$, $f$, or $g$ would occur---the agent cannot predict which---but if $a$ occurred, he would be absolutely certain of it also occurring in a measurement of $H_1^{\rm\small T}$ on the right.  And if $e$ were to occur on the left, then he would be certain of getting $e$ as well in a measurement of $H_5^{\rm\small T}$ on the right.  And similarly with $f$ and $g$, with their implications for $H_8^{\rm\small T}$ and $H_4^{\rm\small T}$.

The wonderful thing to note about (\ref{BurlIves}) is that every letter $a$, $b$, $c$, \ldots, $r$ occurs exactly twice in the collection.  But the EPR criterion of reality (or our Bayesian variant of it) would require exactly one letter to have the truth value TRUE in each column, with the other three having the value FALSE.  In total, nine values of TRUE:  A clean contradiction!  For if every letter occurs exactly twice in the collection, whatever the total number of TRUE values is, it must be an even number.

Something must give.  The quick reaction of most of the quantum foundations community has been to question the causal independence of the two systems under consideration.  But if one gives up on the autonomy of one system from the other---after very explicitly assuming it---this surely amounts to saying that there were never two systems there after all; the very idea of separate systems is a broken concept.  This first raises a minor conundrum:  Why then would the quantum formalism engender us to formulate our description from beginning to end in terms of ${\mathcal H}_4\otimes{\mathcal H}_4$, rather than simply a raw sixteen-dimensional space ${\mathcal H}_{16}$?  Why is that separating symbol $\otimes$, apparently marking some kind of conceptual distinction, always dangling around?

Reaching much deeper however, if one is willing to throw away one's belief in systems' autonomy from each other, why would one ever believe in one's own autonomy?  All stringent reason for it gets lost, and indeed as Einstein warns, what now is the meaning of science?  It is, as Hans Primas wrote somewhere,
\begin{quotation}\small
\noindent a tacit assumption of all engineering sciences that nature can be {\it manipulated\/} and that the initial conditions required by experiments can be brought about by interventions of the world external to the object under investigation.  That is, {\it we assume that the experimenter has a certain freedom of action which is not accounted for by first principles of physics}.  Without this freedom of choice, experiments would be impossible.  Man's free will implies the ability to carry out actions, it constitutes his essence as an actor.  We act under the idea of freedom, but the topic under discussion is neither man's sense of personal freedom as a subjective experience, nor the question whether this idea could be an illusion or not, nor any questions of moral philosophy, but that {\it the framework of experimental science requires the freedom of action as a constitutive though tacit presupposition}.
\end{quotation}
If the left-hand system can manipulate the right-hand system, {\it even when by assumption it cannot}, then who is to say that the right-hand system cannot manipulate the agent himself?\footnote{To put it still differently:  If one is never allowed to assume causal independence between separated systems because of a contradiction in the term, then one can never assume it of oneself either, even with respect to the components of the world one thinks one is manipulating.}  It would be a wackier world than even the one QBism entertains.

But QBism's world is not such a bad world, and some of us find its openness to possibility immensely exciting.  What gives way in this world is simply the EPR criterion of reality:  {\it Both\/} the idea that a probability-one assignment implies there is a pre-existent outcome (property) ``over there'' waiting to be revealed and, baring that, that it must have been ``over here'' pre-existent, waiting to be transferred and then revealed.  The solution lies closer to one of John Wheeler's quips, ``No question? No answer.''  A probability-one assignment lays no necessary claim on what the world {\it is},\footnote{The author believes the opposing opinion on this point is the root of all trouble in arguments claiming to show that Bell inequality violations imply nonlocality, full stop.  Like a clerk at a patent office receiving another proposal for a {\it perpetuum mobile}, the Quantum Bayesian always has to find the singular flawed mechanism that lurks behind the claim---sometimes it is not easy---but it is always there, no matter how sophisticated the argument. One finds oneself thankful for the very clearest papers on the subject, for they practically lay the point on a tray.  A good example is Travis Norsen's presentation \cite{Norsen06}, where it is written:%
\begin{quotation}\footnotesize
``\ldots a statement of the form \ldots\ implies one of the form
$$
P(A = -1|\hat{n}_1, \lambda) = 1
$$
since the outcomes are bivalent: if, for a given [hidden-variable state] $\lambda$ and a given measurement direction [$\hat{n}_1$], a certain outcome is (according to some theory) {\it impossible}, then, since there are only two possible outcomes, the opposite outcome is {\it required}. \ldots\ This suggests a shorthand notation in which we substitute \ldots\ the simpler statement
$$
A(\hat{n}_1, \lambda) = -1
$$

The reader may worry that we are here violating Wheeler's famous statement of the orthodox quantum philosophy \ldots\ i.e., it is invalid to attribute particular outcomes to experiments which haven't, in fact, been performed. This worry is partly justified. We are not, however, asserting
that an un-performed measurement has an actual, particular outcome; this would be literal nonsense, and is the grain of truth in Wheeler's dictum. Strictly speaking, our statement isn't even {\it about\/} Alice's measurement -- it is about the state $\lambda$ and the theory in which that state
assignment is embedded. The real meaning of [the last equation above] is simply this: for the state $\lambda$, the theory in question assigns
unit probability to the outcome $A = -1$ under the condition that Alice measures along direction $\hat{n}_1$. The theory must attribute sufficient structure to $\lambda$ (and possess the necessary dynamical laws) such that, should Alice choose to measure along $\hat{n}_1$, the outcome $A = -1$ is guaranteed. In this sense, we may say that the theory in question {\it encodes\/} the outcome $A = -1$ (for measurement along $\hat{n}_1$) in the state $\lambda$.''\end{quotation}} but what the agent using it believes with all his heart and soul.  In the case of our present example, what the agent believes is that if an outcome $b$ {\it came about\/} as a result of his action $H$ on the left-hand system, an outcome $b$ {\it would come about\/} if he were to perform the action $H^{\rm\small T}$ on the right-hand system.  But if he does not walk over to the right-hand system and take the action, there is no good sense in which the outcome (or property) $b$ is already there.

At the instigation of a quantum measurement, something new comes into the world that was not there before; and that is about as clear an instance of {\it creation\/} as one can imagine.  Sometimes one will have no strong beliefs for what will result from the creation (as with the measurement of $H$), and sometimes one will have very strong beliefs (as with the subsequent measurement of $H^{\rm\small T}\,$), but a free creation of nature it remains.

\section{Hilbert-Space Dimension as a Universal Capacity}

\label{HSpaceDim}

\begin{flushright}
\baselineskip=13pt
\parbox{2.8in}{\baselineskip=13pt\small
It is entirely possible to conceive of a world composed of individual atoms, each as different from one another as one organism is from the next.}
\medskip\\
\small --- John Dupr\'e
\end{flushright}

A common accusation heard by the Quantum Bayesian is that the view leads straight away to solipsism, ``the belief that all reality is just one's  imagining of reality, and that one's self is the only thing that exists.''\footnote{This is the definition of {\sl The American Heritage New Dictionary of Cultural Literacy}, Third Edition (2005).  {\sl Encyclopedia Brittanica\/} (2008) expands, ``in philosophy \ldots\ the extreme form of subjective idealism that denies that the human mind has any valid ground for believing in the existence of anything but itself. The British idealist F.~H. Bradley, in {\sl Appearance and Reality\/} (1897), characterized the solipsistic view as follows: `I cannot transcend experience, and experience is my experience. From this it follows that nothing beyond myself exists; for what is experience is its (the self's) states.'$\,$''}  The accusation goes that, if a quantum state $|\psi\rangle$ only represents the degrees of belief held by some agent---say, the one portrayed in Figure 1---then the agent's beliefs must be the source of the universe.  The universe could not exist without him: This being such a ridiculous idea, QBism is dismissed out of hand, {\it reductio ad absurdum}.  It is so hard for the QBist to understand how anyone could think this (it being the antithesis of everything in his worldview) that a little of our own Latin comes to mind: {\it non sequitur}.  See Figure 5.

A fairer-minded assessment is that the accusation springs from our opponents ``hearing'' much of what we do say, but interpreting it in terms drawn from a particular conception of what physical theories {\it always ought to be}:  Attempts to directly represent (map, picture, copy, correspond to, correlate with) the {\it universe}---with ``universe'' here thought of in totality as a pre-existing, static system; an unchanging, monistic something that just {\it is}.  From such a ``representationalist'' point of view, {\it if\/} {\bf a)} quantum theory is a proper physical theory, {\bf b)} its essential theoretical objects are quantum states, and {\bf c)} quantum states are states of belief, {\it then\/} the universe that ``just is'' corresponds to a state of belief.  Solipsism on a stick, one might say.\footnote{See {\tt http://www.urbandictionary.com/define.php?term =on+a+stick\/} if you have any doubt of the meaning.}

\begin{figure}
\begin{center}
\includegraphics[height=2.0in]{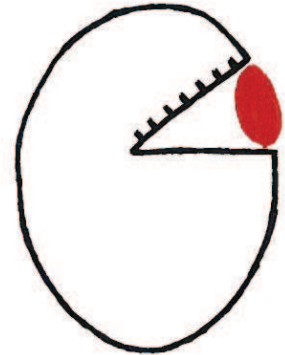}
\bigskip\caption{\protect\small {\bf Sarcasm.}~In a lecture bottlenecked by repeated accusations of Quantum Bayesianism's solipsism, the author sometimes uses the following technique to move things along. Referring to the previous Figure 1, he asks the stubborn accuser, ``What about this diagram do you {\it not\/} get?  It shows an agent and a physical system external to him.  It says that a quantum state is a {\it state of belief\/} about what will come about as a consequence of his actions upon the system.   The quantum state is not a state of nature, but so what?  There is an agent with his belief; there is a system that is not part of him; and there is something that really, eventually comes about---it is called {\it the outcome}.  No agent, no outcome for sure, but that's not solipsism:  For, no system, no outcome either!  A quantum measurement without an external system participating would be like the sound of one hand clapping, a Zen koan.  If we were really expressing solipsism, wouldn't a diagram like the one above be more appropriate?  A big eyeball surveying nothing.  Now there's really no external system and nothing to act upon. {\it That's\/} solipsism.''}
\end{center}
\end{figure}

Quantum Bayesianism sidesteps the poisoned dart, as the previous sections have tried to convey, by asserting that quantum theory is just not a physical theory in the sense the accusers want it to be.  Rather it is an addition to personal, Bayesian, normative probability theory.  Its normative rules for connecting probabilities (personal judgments) were developed in light of the {\it character of the world}, but there is no sense in which the quantum state itself represents (pictures, copies, corresponds to, correlates with) a part or a whole of the external world, much less a world that {\it just is}.  In fact the very character of the theory seems to point to the inadequacy of the representationalist program when attempted on the particular world we live in.

There are no lofty philosophical arguments here that re\-presentationalism must be wrong always and in all possible worlds (perhaps because of some internal inconsistency\footnote{As \cite{Rorty91} might try to argue.}).  Representationalism may well be true in this or that setting---we take no stand on the matter.  We only know that for nearly 90 years quantum theory has been actively resistant to representationalist efforts on {\it its\/} behalf.  This suggests that it might be worth exploring some philosophies upon which physics rarely sets foot.  Physics of course should never be constrained by any one philosophy (history shows it nearly always lethal), but it does not hurt to get ideas and insights from every source one can.  If one were to sweep the philosophical literature for schools of thought representative of what QBism actually is about, it is not solipsism one will find, but nonreductionism \cite{Dupre93,Cartwright99}, (radical) metaphysical pluralism \cite{James96a,Wahl25}, empiricism \cite{James40,James96b}, indeterminism and meliorism\footnote{Strictly speaking, meliorism is the doctrine ``that humans can, through their interference with processes that would otherwise be natural, produce an outcome which is an improvement over the aforementioned natural one.''  But we would be reluctant to take a stand on what ``improvement'' really means.  So said, all we mean in the present essay by meliorism is that the world before the agent is malleable to some extent---that his actions really can change it.  Adam said to God, ``I want the ability to write messages onto the world.''  God replied, ``You ask much of me.  If you want to write upon the world, it cannot be so rigid a thing as I had originally intended.  The world would have to have some malleability, with enough looseness for you to write upon its properties.  It will make your world more unpredictable than it would have been---I may not be able to warn you about impending dangers like droughts and hurricanes as effectively as I could have---but I can make it such if you want.'' And with that Adam brought all host of uncertainties to his life, but he gained a world where his deeds and actions mattered.} \cite{James1884}, and above all pragmatism \cite{Menand01,Thayer81}.

A form of nonreductionism can already be seen in play in our answer to whether the notion of agent should be derivable from the quantum formalism itself.  We say that it cannot be and it should not be, and to believe otherwise is to misunderstand the subject matter of quantum theory.  But nonreductionism also goes hand in hand with the idea that there is real particularity and ``interiority'' in the world.  Think again of the ``I-I-me-me mine'' feature that shields QBism from inconsistency in the ``Wigner's friend'' scenario.
When Wigner turns his back to his friend's interaction with the system, that piece of reality---Bohr might call it a ``phenomenon''\footnote{With the mention of Bohr's beloved ``phenomenon,'' the author atones for the sin explained in Footnote \ref{FiddleFotts}.  That said, we stress the word {\it might}.  Here is the way H.~J. Folse \cite{Folse87b} explains Bohr's conception of the word:
\begin{quotation}\footnotesize
Bohr repeatedly stressed that the break between classical and quantum physics, resides in the fact that `Planck's discovery of the {\it elementary quantum of action\/} \ldots\ revealed a feature of {\it wholeness\/} inherent in atomic processes, going far be\-yond the ancient idea of the limited divisibility of matter.'  The consequence of adopting the quantization of action in physics is that in the fine structure of the physical world it is these {\it causal pro\-cesses of interaction}, not bits of material substances, which are `atomized'---i.e.\ made `indivisible.'  The `atomicity of an interaction' implies that the description of an `observation' of a microsystem which must subdivide the whole interaction into separate observed and observing systems is a conceptual idealization or `abstraction' employed for interpreting the observation, but not a `picture' of an objective course of events.  On Bohr's concept of reality we in\-di\-vid\-uate the parts of the world not {\it qua\/} spatio-temporally `sep\-arable' physical systems possessing properties apart from any interaction, but instead {\it qua\/} `individual' interactions each of which is a whole phenomenon the description and prediction of which must be the goal of a successful physical theory.\end{quotation}}---is hermetically sealed from him.  It has an inside, a vitality that he takes no part in until he again interacts with one or both relevant pieces of it.  {\it With respect to Wigner}, it is a bit like a universe unto itself.\footnote{Would it be a universe unto itself with respect to you the reader?  I cannot say:  It would depend upon {\it your\/} quantum state for the friend and system, and that might have nothing to do with Wigner's.  But it depends upon even more than that. For you might surmise the friend to be in contact with physical systems Wigner never even dreamed of.  That is, you may not assign the same relevant systems as Wigner---your assignment might not be bipartite, but multipartite.  When one deems quantum states personalist Bayesian assignments, all these options must be taken seriously and savored for the lessons they teach.}

If one seeks the essence of indeterminism in quantum mechanics, there may be no example more directly illustrative of it than ``Wigner's friend.''  For it expresses to a tee William James's notion of indeterminism \cite{James1884}:
\begin{quotation}\small
[Chance] is a purely negative and relative term, giving us no
information about that of which it is predicated, except that it
happens to be disconnected with something else---not controlled,
secured, or necessitated by other things in advance of its own actual
presence. \ldots\ What I say is that it tells us
nothing about what a thing may be in itself to call it ``chance.'' \ldots\
All you mean by calling it ``chance'' is that this is not guaranteed,
that it may also fall out otherwise. For the system of other things
has no positive hold on the chance-thing. Its origin is in a certain
fashion negative: it escapes, and says, Hands off!\ coming, when it
comes, as a free gift, or not at all.

This negativeness, however, and this opacity of the chance-thing when
thus considered {\it ab extra}, or from the point of view of previous
things or distant things, do not preclude its having any amount of
positiveness and luminosity from within, and at its own place and
moment. All that its chance-character asserts about it is that there
is something in it really of its own, something that is not the
unconditional property of the whole. If the whole wants this
property, the whole must wait till it can get it, if it be a matter
of chance. That the universe may actually be a sort of joint-stock
society of this sort, in which the sharers have both limited
liabilities and limited powers, is of course a simple and conceivable
notion.
\end{quotation}
And once again \cite{JamesOSH},
\begin{quotation}\small
Why may not the world be a sort of republican banquet of this sort, where all the qualities of being respect one another's personal sacredness, yet sit at the common table of space and time?

To me this view seems deeply probable.  Things cohere, but the act of cohesion itself implies but few conditions, and leaves the rest of their qualifications indeterminate.  As the first three notes of a tune comport many endings, all melodious, but the tune is not named till a particular ending has actually come,---so the parts actually known of the universe may comport many ideally possible complements. But as the facts are not the complements, so the knowledge of the one is not the knowledge of the other in anything but the few necessary elements of which all must partake in order to be together at all. Why, if one act of knowledge could from one point take in the total perspective, with all mere possibilities abolished, should there ever have been anything more than that act? Why duplicate it by the tedious unrolling, inch by inch, of the foredone reality? No answer seems possible. On the other hand, if we stipulate only a partial community of partially independent powers, we see perfectly why no one part controls the whole view, but each detail must come and be actually given, before, in any special sense, it can be said to be determined at all.  This is the moral view, the view that gives to other powers the same freedom it would have itself.
\end{quotation}

The train of (still loose, but slowly firming) logic back to QBism is this.  If James and our analysis of ``Wigner's friend'' are right, the universe is not {\it one\/} in a very rigid sense, but rather more truly a pluriverse.\footnote{The term ``pluriverse'' is again a Jamesian one.  He used it interchangeably with the word ``multiverse,'' which he also invented.  Unfortunately the latter has been coopted by the Everettian movement for their own---in the end monistic---purposes:  ``The world is one; it {\it is\/} the deterministically evolving universal quantum state, the `multiverse'.'' Too bad.  Multiverse is a tempting word, but we stick with pluriverse to avoid any confusion with the Everettian usage.}  To get some sense of what this can mean, it is useful to start by thinking about what it is not.  A good example can be found by taking a solution to the vacuum Maxwell equations in some extended region of spacetime.  Focus on a compact subregion and try to conceptually delete the solution within it, reconstructing it with some new set of values.  It can't be done.  The fields outside the region (including the boundary) uniquely determine the fields inside it.  The interior of the region has no identity but that dictated by the rest of the world---it has no ``interiority'' of its own.  The pluriverse conception says we'll have none of that.  And so, for any agent immersed in this world there will always be uncertainty for what will happen upon his encounters with it.  To wit, where there is uncertainty there should be Bayesian probabilities, and so on and so on until much of the story we have already told.

What all this hints is that for QBism the proper way to think of our world is as the empiricist or radical metaphysical pluralist does.  Let us launch into making this clearer, for that process more than anything will explain how QBism hopes to interpret Hilbert-space dimension.

The metaphysics of empiricism can be put like this.  Every\-thing experienced, everything experienceable\footnote{That is, every piece of the universe had better be hard-wired for the contingency that an agent might experience it somewhere, somehow, no matter how long and drawn out the ultimate chain might be to such a potential experience.  Does this mean even ``elementary'' physical events just after the Big Bang must make use of {\it concepts\/} that, to the reductionist mind, ought to be 15 billion years removed down the evolutionary chain?  You bet it does.  But a nonreductionist metaphysic need make no apology for this---such things are in the very idea.  John Wheeler's great smoky dragon \cite{Miller84} comes into the world biting its own tail.%

W.~K. Wootters tells a lovely story of an encounter he had several years ago of with his young son Nate.  Nate said, ``I wish I could make this flower move with my mind.''  Wootters reached out and pushed the flower, saying, ``You can. You do it like this.''  From the perspective here, this is an example of an interaction between two nonreductionist realms.  Each realm influences the other as its turn comes.  There is a kind of reciprocality in this, an action-reaction principle, that most reductionist visions of the world would find obscene.},
has no less an ontological status than anything else.  You tell me of your experience, and I will say it is real, even a distinguished part of reality.  A child awakens in the middle of the night frightened that there is a monster under her bed, one soon to reach up and steal her arm---that {\it we-would-call-imaginary\/} experience\footnote{There is indeed no doubt that it should be called imaginary!  That, however, is a statement about the experience's meaning and interpretation, not its existence.  The ex\-pe\-rience {\it as it is\/} exists, period. It is what it is.  Like the biblical burning bush, each ex\-pe\-rience declares, ``I am that I am.''  Most likely in the present example, the experience will be a little piece of the universe isolated, on its own, and of no great consequence.  But one never knows until all future plays out.  Some lucky dreams have built nations.  Maybe the same is true of some lucky Higgs-boson events.  Most though, surely, will be of the more minor fabric of existence.} has no less a hold on onticity than a Higgs-boson detection event would if it were to occur at the fully operational LHC.  They are of equal status from this point of view---they are equal elements in the filling out and making of reality.  This is because the world of the empiricist is not a sparse world like the world of Democritus ({\it nothing but\/} atom and void) or Einstein ({\it nothing but\/} unchanging spacetime manifold equipped with this or that field), but a world overflowingly full of variety---a world whose details are beyond anything grammatical (rule-bound) expression can articulate.

Yet this is no statement that physics should give up, or that physics has no real role in coming to grips with the world.  It is only a statement that physics should better understand its function.  What is being aimed for here finds its crispest, clearest contrast in a statement Richard Feynman once made \cite{Feynman65}:
\begin{quotation}\small
If, in some cataclysm, all of scientific knowledge were to be destroyed, and only one sentence passed on to the next generation of creatures, what statement would contain the most information in the fewest words?  I believe it is the atomic hypothesis (or the atomic fact) that all things are made of atoms---little particles that move around in perpetual motion, attracting each other when they are a little distance apart, but repelling upon being squeezed into one another. \ldots

Everything is made of atoms.  That is the key hypothesis.
\end{quotation}
The issue for QBism hangs on the imagery that usually lies behind the phrase ``everything is made of.''  William James called it the great original sin of the rationalistic mind \cite{James97}:
\begin{quotation}\small
Let me give the name of `vicious abstractionism' to a way of using concepts which may be thus described: We conceive a concrete situation by singling out some salient or important feature in it, and classing it under that; then, instead of adding to its previous characters all the positive consequences which the new way of conceiving it may bring, we proceed to use our concept privatively; reducing the originally rich phenomenon to the naked suggestions of that name abstractly taken, treating it as a case of `nothing but' that, concept, and acting as if all the other characters from out of which the concept is abstracted were expunged. Abstraction, functioning in this way, becomes a means of arrest far more than a means of advance in thought. It mutilates things; it creates difficulties and finds impossibilities; and more than half the trouble that metaphysicians and logicians give themselves over the paradoxes and dialectic puzzles of the universe may, I am convinced, be traced to this relatively simple source. {\it The viciously privative employment of abstract characters and class names\/} is, I am persuaded, one of the great original sins of the rationalistic
mind.
\end{quotation}
What is being realized through QBism's peculiar way of looking at things is that physics {\it actually can be done\/} without any accompanying vicious abstractionism.  You do physics as you have always done it, but you throw away the idea ``everything is made of [Essence X]'' before even starting.

Physics---in the right mindset---is not about identifying the bricks with which nature is made, but about identifying what is {\it common to\/} the largest range of phenomena it can get its hands on.  The idea is not difficult once one gets used to thinking in these terms.  Carbon?  The old answer would go that it is {\it nothing but\/} a building block that combines with other elements according to the following rules, blah, blah, blah.  The new answer is that carbon is a {\it characteristic\/} common to diamonds, pencil leads, deoxyribonucleic acid, burnt pancakes, the space between stars, the emissions of Ford pick-up trucks, and so on---the list is as unending as the world is itself.  For, carbon is also a characteristic common to this diamond and this diamond and this diamond and this.  But a flawless diamond and a purified zirconium crystal, no matter how carefully crafted, have no such characteristic in common:  Carbon is not a {\it universal\/} characteristic of all phenomena.  The aim of physics is to find characteristics that apply to as much of the world in its varied fullness as possible.  However, those common characteristics are hardly what the world is made of---the world instead is made of this and this and this.  The world is constructed of every particular there is and every way of carving up every particular there is.

An unparalleled example of how physics operates in such a world can be found by looking to Newton's law of universal gravitation.  What did Newton really find?  Would he be considered a great physicist in this day when every news magazine presents the most cherished goal of physics to be a Theory of Everything?  For the law of universal gravitation is hardly that!  Instead, it {\it merely\/} says that every body in the universe tries to accelerate every other body toward itself at a rate proportional to its own mass and inversely proportional to the squared distance between them.  Beyond that, the law says nothing else particular of ob\-jects, and it would have been a rare thinker in Newton's time, if any at all, who would have imagined that all the complexities of the world could be derived from that limited law.  Yet there is no doubt that Newton was one of the greatest physicists of all time.  He did not give a theory of everything, but a Theory of One Aspect of Everything.  And only the tiniest fraction of physicists of any variety, much less the TOE-seeking variety, have ever worn a badge of that more modest kind.  It is as H.~C. von Baeyer wrote in one of his books \cite{Baeyer09},
\begin{quote}\small
Great revolutionaries don't stop at half measures if they can go all the way.  For Newton this meant an almost unimaginable widening of the scope of his new-found law.  Not only Earth, Sun, and planets attract objects in their vicinity, he conjectured, but all objects, no matter how large or small, attract all other objects, no matter how far distant.  It was a proposition of almost reckless boldness, and it changed the way we perceive the world.
\end{quote}
Finding a theory of ``merely'' one aspect of everything is hardly something to be ashamed of:  It is the loftiest achievement physics can have in a living, breathing non\-re\-duc\-tionist world.

Which leads us back to Hilbert space.  Quantum theory---that user's manual for decision-making agents immersed in a world of {\it some\/} yet to be fully identified character---makes a statement about the world to the extent that it identifies a quality common to all the world's pieces.  QBism says the quantum state is not one of those qualities.  But of Hilbert spaces themselves, particularly their distinguishing characteristic one from the other, {\it dimension},\footnote{Hardy \cite{Hardy01a,Hardy01b} and Daki\'c and Brukner \cite{Dakic09} are examples of foundational efforts that also emphasize this quantum analog to what E\"otv\"os tested on platinum and copper \cite{Fuchs04b}.  Hardy put it this way in one of his axioms, ``There exist systems for which $N = 1, 2, \cdots$, and, furthermore, all systems of dimension $N$, or systems of higher dimension but where the state is constrained to an $N$ dimensional subspace, have the same properties.''} QBism car\-ries no such grudge.  Dimension is something one posits for a body or a piece of the world, much like one posits a mass for it in the Newtonian theory.  Dimension is something a body holds all by itself, regardless of what an agent thinks of it.

That this is so can be seen already from reasons internal to the theory.  Just think of all the arguments rounded up for making the case that quantum states should be interpreted as of the character of Bayesian degrees of belief.  None of these work for Hilbert-space dimension.  Take one example, an old favorite---Einstein's argument about conditioning quantum states from afar.  In Section \ref{BellTheorem} of this paper we repeated the argument verbatim, but it is relevant to note that before Einstein could write down his $\psi_{12}$, he would have had to associate some Hilbert spaces ${\mathcal H}_1$ and ${\mathcal H}_2$ with $S_1$ and $S_2$ and take their tensor product ${\mathcal H}_1\otimes{\mathcal H}_2$.  Suppose the dimensionalities of these spaces to be $d_1$ and $d_2$, respectively.  The question is, is there anything similar to Einstein's argument for changing the value of $d_2$ from a distance?  There isn't.  $\psi_2$ may be forced into this or that subspace by choosing the appropriate measurement on $S_1$, but there is no question of the whole Hilbert space ${\mathcal H}_2$ remaining intact.  When it is time to measure $S_2$ itself, one will still have the full arsenal of quantum measurements appropriate to a Hilbert space of dimension $d_2$ to choose from---none of those fall by the wayside.  In Einstein's terms, $d_2$ is part of the ``real factual situation'' of $S_2$.\footnote{Most recently techniques have started to become available to ``test'' the supposition of a dimension against one's broader mesh of beliefs; see \cite{Brunner08,Wehner08,Wolf09}.}

The claim here is that quantum mechanics, when it came into existence, implicitly recognized a previously unnoticed capacity inherent in all matter---call it {\it quantum dimension}.  In one manifestation, it is the fuel upon which quantum computation runs \cite{Fuchs04b,BlumeKohout02}.  In another it is the raw irritability of a quantum system to being eavesdropped upon \cite{Fuchs03,Cerf02}.  In Eq.~(\ref{ScoobyDoo}) it was a measure of deviation from the Law of Total Probability induced by counterfactual thinking.  And in a farther-fetched scenario to which we will come back, its logarithm {\it might\/} just manifest itself as the squared gravitational mass of a Schwarzschild black hole \cite{Horowitz04,Gottesman04}.

When quantum mechanics was discovered, something was {\it added\/} to matter in our conception of it.  Think of the apple that inspired Newton to his law.  With its discovery the color, taste, and texture of the apple didn't disappear; the law of universal gravitation didn't reduce the apple privatively to {\it just\/} gravitational mass. Instead, the apple was at least everything it was before, but afterward even more---for instance, it became known to have something in common with the moon.  A modern-day Cavendish would be able to literally measure the fur\-ther attraction an apple imparts to a child already hungry to pick it from the tree.  So similarly with Hil\-bert-space dimension.  Those diamonds we have already used to illustrate the idea of nonreductionism, in very careful conditions, could be used as components in a quantum computer \cite{Prawer08}.  Diamonds have among their many properties something not envisioned before quantum mechanics---that they could be a source of relatively accessible Hilbert space dimension and as such have this much in common with any number of other proposed implementations of quantum computing.  Diamonds not only have something in common with the moon, but now with the ion-trap quantum-computer prototypes around the world.

Diamondness is not something to be derived from quantum mechanics.  It is that quantum mechanics is something we {\it add\/} to the repertoire of things we already say of diamonds, to the things we do with them and the ways we admire them.  This is a very powerful realization:  For diamonds already valuable, become ever more so as their qualities compound.  And saying more of them, not less of them as is the goal of all reductionism, has the power to suggest all kinds of variations on the theme.  For instance, thinking in quantum mechanical terms might suggest a technique for making ``purer diamonds.''  Though to an empiricist this phrase means not at all what it means to a reductionist.  It means that these similar things called diamonds can suggest exotic variations of the original objects with various pinpointed properties this way or that.  Purer diamond is not {\it more\/} of what it already was in nature.  It is a new species, with traits of its parents to be sure, but nonetheless stand-alone, like a new breed of dog.

To put it still differently, and now in the metaphor of music, a jazz musician might declare that a tune once heard thereafter plays its most crucial role as a substrate for something new. It is the fleeting solid ground upon which something new can be born.  The seven tracks titled {\sl Salt Peanuts\/} in my mp3 player\footnote{Charlie Parker, Dizzy Gillespie, Charlie Parker, Charlie Parker, Charlie Parker, Joshua Redman, Miles Davis Quintet.} are moments of novelty in the universe never to be recreated.  So of diamonds, and so of all this quantum world.  Or at least that is the path QBism seems to indicate.\footnote{A nice {\it logical\/} argument for this can be found in \cite{Ojima92}.}

To the reductionist, of course, this seems exactly backwards.  But then, it is the reductionist who must live with a seemingly infinite supply of conundrums arising from quantum mechanics.  It is the reductionist who must live in a state of arrest, rather than moving on to the next stage of physics.  Take a problem that has been a large theme of the quantum foundations meetings for the last 30 years.  To put it in a commonly heard question, ``Why does the world look classical if it actually operates according to quantum mechanics?''  The touted mystery is that we never ``see'' quantum superposition and entanglement in our everyday experience.\footnote{Of course, to a group of personalist Bayesians that's like asking, ``Which of you has ever seen a probability distribution?''  Not a one will say yes.  Probabilities in personalist Bayes\-ian\-ism are not the sorts of things that can be seen; they are the things that are thought.  It is {\it events\/} that are seen.  But let us drop the matter for the moment.}

The real issue is this.  The expectation of the quantum-to-classical transitionists\footnote{See \cite{Schlosshauer07,Schlosshauer08} for particularly clear discussions of the subject.} is that quantum theory is at the bottom of things, and ``the classical world of our experience'' is something to be derived out of it.  QBism says ``No.  Experience is neither classical nor quan\-tum.  Experience is experience with a richness that classical physics of any variety could not remotely grasp.''  Quantum mechanics is something put on top of raw, unreflected experience.  It is additive to it, suggesting wholly new types of experience, while never invalidating the old.  To the question, ``Why has no one ever {\it seen\/} superposition or entanglement in diamond before?,'' the QBist replies:  It is simply because before recent technologies and very controlled conditions, as well as lots of refined analysis and thinking, no one had ever mustered a mesh of beliefs relevant to such a range of interactions (factual and counterfactual) with diamonds.  No one had ever been in a position to adopt the extra normative constraints required by the Born Rule.  For QBism, it is not the emergence of classicality that needs to be explained, but the emergence of our new ways of manipulating, controlling, and interacting with matter that do.

In this sense, QBism declares the quantum-to-classical research program unnecessary (and actually obstructive\footnote{Without an ontic understanding of quantum states, quantum operations, and unitary time evolutions---all of which QBism rejects, see Footnote \ref{Quasiclosure} and Refs.~\cite{Fuchs02,Fuchs09a,Leifer06}---how can the project even get off the ground?  As one can ask of the Big Bang, ``What banged?,'' the QBist must ask, ``In those days of the world before agents using quantum theory, what decohered?''}) in a way not so dissimilar to the way Bohr's 1913 model of the hydrogen atom declared another research program unnecessary (and actually obstructive).\footnote{All is not lost, however, for the scores of decoherentists this policy would unforgivingly unemploy.  For it only suggests that they redirect their work to the opposite direction.  The thing that needs insight is not the quantum-to-classical transition, but the classical-to-quantum!  The burning question for the QBist is how to model in Hilbert-space terms the common sorts of measurements we perform just by opening our eyes, cupping our ears, and extending our fingers.%

Take a professional baseball player watching a ball fly toward him:  He puts his whole life into when and how he should swing his bat.  But what does this mean in terms of the immense Hilbert space a quantum theoretical description would associate with the ball?  Surely the player has an intuitive sense of both the instantaneous position and instantaneous momentum of the baseball before he lays his swing into it---that's what ``keeping his eye on the ball'' means.  Indeed it is from this intuition that Newton was able to lay down his laws of classical mechanics.   Yet, what can it mean to say this given quantum theory's prohibition of simultaneously measuring complementary observables?  It means that whatever the baseball player is measuring, it ain't that---it ain't position {\it and\/} momentum as usually written in operator terms.  Instead, a quantum model of what he is doing would be some interesting, far-from-extremal {\it single\/} POVM---perhaps even one that takes into account some information that does not properly live within the formal structure of quantum theory (the larger arena that Howard Barnum calls ``meaty quantum physics'' \cite{Barnum10}). For instance, that an eigenvector $|i\rangle$ of some Hermitian operator, though identically orthogonal to fellow eigenvectors $|j\rangle$ and $|k\rangle$ in the Hilbert-space sense, might be {\it closer\/} in meaning to $|k\rangle$ than to $|j\rangle$ for some issue at hand.%

So the question becomes how to take a given common-day measurement procedure and {\it add to it\/} a consistent quantum description?  The original procedure was stand alone---it can live without a quantum description of it---but if one wants to move it to a new level or new direction, having added a consistent quantum description will be most helpful to those ends.  Work along these lines is nascent, but already some excellent examples exist.  See \cite{Kofler08}.  Of course, unconsciously it is what has been happening since the founding days of quantum mechanics.}  Bohr's great achieve\-ment above all the other physicists of his day was in being the first to say, ``Enough!  I shall not give a mech\-anistic explanation for these spectra we see.  Here is a way to think of them with no mechanism.''  The important question is how matter can be coaxed to do new things.  It is in the ways the world yields to our desires, and the ways it refuses to, that we learn the depths of its character.
\begin{verse}\small
I give you an object of this much gravitational mass.  What can you do with it?  What can you not?  And when you are not about, what does it cause?
\\
I give you an object of this much quantum dim\-ension.  What can you do with it?  What can you not?  And when you are not about, what does it cause?
\end{verse}

If taken seriously what do these questions imply by their very existence?  That they should have meaningful answers!  Here is one example.  A knee-jerk reaction in many physicists upon hearing these things is to declare that dimension as a capacity collapses to a triviality as soon as it is spoken.  ``All real-world systems possess infinite-dimensional Hilbert spaces.  And it doesn't take quantum field theory to be completely correct to make that true; a simple one-dimensional harmonic oscillator will do.  It has an infinite-dimensional Hilbert space.''  But maybe not.  Maybe no real-world quantum system has that much oomph.  Just as one can treat the Earth's inertial mass as infinite for many a freshman mechanics problem, or a heat bath as infinite for many a thermodynamical one, maybe this is all that has ever been going on with infinite-dimensional Hilbert spaces.  It is a useful artifice when a problem can be economically handled with a differential equation.  (Ask Schr\"odinger.)

And with this, we come to nearly the farthest edge of QBism.  It is the beginning of a place where quantum mechanics must step past itself.  To make quantum dimension meaningful in ontic terms, as a quality common to all physical objects, is to say it should be finite---going up, going down from this object to the next, but always finite.  Every region of space where electromagnetism can propagate, finite.  Every region of space where there is a gravitational ``field,'' finite.

It means that despite its humble roots in nonrelativistic quantum mechanics, there is something already cosmological about QBism.  It tinkers with spacetime, saying that in every ``hole'' (every bounded region) there is an interiority not given by the rest of the universe and a common quality called dimension.  It says that there is probably something right about the holographic principles arising from other reaches of physics \cite{Verlinde10}.  But also Quantum Bayesianism, recognizing entropy as a personal concept (entropy is a function of probability), would suspect that it is not an entropy bound that arises from these principles.  Would it be a dimension bound?  And why so, if there were not a new equivalence principle lurking around the corner \cite{Fuchs04b,Fuchs10b}?

\section{Quantum Cosmology from the Inside}

Let us, however, step back from that farthest edge for a moment and discuss cosmology as it is presently construed before taking that final leap!

\begin{figure}
\begin{center}
\includegraphics[height=3.6in]{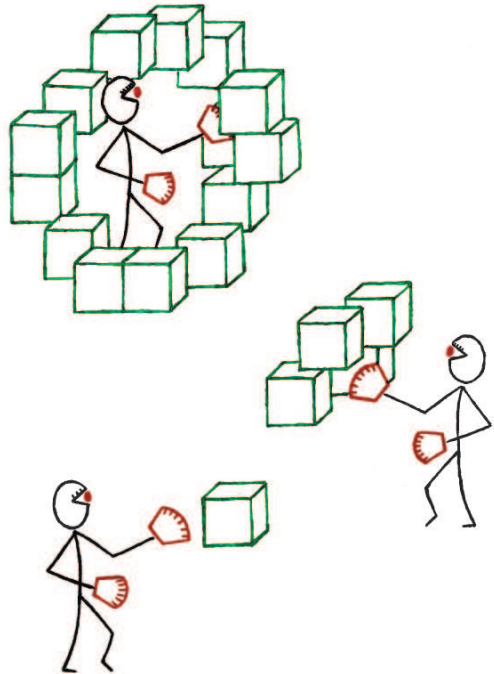}
\bigskip\caption{\protect\small {\bf Quantum Cosmology from the Inside.}~The agent in Figure 1 can consider measurements on ever larger systems. There is nothing in quantum mechanics to bar the systems considered from being larger and larger, to the point of eventually surrounding the agent.  Pushed far enough, this is quantum cosmology!  Why all this insistence on thinking that ``an agent must be outside the system he measures'' in the cosmological context should mean ``outside the physical universe itself''?  It means outside the system of interest, and that is the large-scale universe. Nor is there any issue of self-reference at hand.  One would be hard pressed to find a cosmologist who wants to include his beliefs about how the beats of his heart correlate with the sidereal cycles in his quantum-state assignment for the external universe.  The symbol $|\,\Psi_{\rm \small universe}\,\rangle$ refers to the green boxes alone.}
\end{center}
\end{figure}

Sometimes it is claimed that a point of view about quantum theory like QBism's would make the enquiries of quantum cosmology impossible.  For instance, David Deutsch once put it like this \cite{Deutsch86}:
\begin{quotation}
\small The best physical reason for adopting the Everett interpretation lies in quantum cosmology.  There one tries to apply quantum theory to the universe as a whole, considering the universe as a dynamical object starting with a big bang, evolving to form galaxies and so on. Then when one tries, for example by looking in a textbook, to ask what the symbols in the quantum theory mean, how does one use the wave function of the universe and the other mathematical objects that quantum theory employs to describe reality?  One reads there, `The meaning of these mathematical objects is as follows:  first consider an observer outside the quantum system under consideration \ldots.' And immediately one has to stop short.  Postulating an outside observer is all very well when we're talking about a laboratory:  we can imagine an observer sitting outside the experimental apparatus looking at it, but when the experimental apparatus---the object being described by quantum theory---is the entire universe, it's logically inconsistent to imagine an observer sitting outside it. Therefore the standard interpretation fails.  It fails completely to describe quantum cosmology.  Even if we knew how to write down the theory of quantum cosmology, which is quite hard incidentally, we literally wouldn't know what the symbols meant under any interpretation other than the Everett interpretation.
\end{quotation}
But this is nonsense.  It is not hard to imagine how to measure the universe as a whole:  You simply live in it.

What are the typical observables and predictables of cosmology?  The Hubble constant, the cosmological constant, the degree of inhomogeneity of the cosmic microwave background radiation, total baryon number in this or that era of the universe, perhaps others.  To do quantum cosmology is to ask how an application of quantum mechanics can be made with regard to these quantities.  For the Quantum Bayesian quantum theory would be used {\it as it always is}:  As a normative calculus of consistency for all probability assignments concerned.  Quantum theory advises an agent to make all his probability assignments derivable from a single quantum state.  Write it like this if you wish:
\begin{equation}
|\,\Psi_{\rm \small universe}\,\rangle
\label{Muggle}
\end{equation}
why not?\footnote{Well, there is a reason why not. One doesn't even write down a pure quantum state for laser light when its phase is unknown; a mixed state is more appropriate \cite{vanEnk02}.  It is hard to imagine why one would write down a pure state for the large-scale universe.  Who would have beliefs that strict of it?  Be that as it may, a pure state is certainly allowed in principle.  Even people with the most unreasonable of initial beliefs (from one's own perspective) want to gamble consistently.} We are swimming in this ocean called the universe, and we have to do physics from inside of it.  But then all the rest of the universe is outside each of us.  Eq.~(\ref{Muggle}) represent an agent's catalog of beliefs for the relevant things outside.

The only point here is that QBism has every bit as much right to do cosmology as any other crazy interpretation of quantum mechanics.  The only difference is that QBism does it from the inside.

More exciting is the possibility that once it does all that (its own version of what the other interpretations might have done), its power may not be exhausted.  For, noting how the Big Bang itself is a moment of creation with some resemblance to every individual quantum measurement, one starts to wonder whether even it ``might be on the inside.''  Certainly QBism has creation going on all the time and everywhere; quantum measurement is just about an agent hitching a ride and partaking in that ubiquitous process.

At the end of a long article it doesn't hurt to speculate.  We let William James and John Archibald Wheeler do the work for us.  First more sweepingly \cite{James22},
\begin{quotation}\small
Our acts, our turning-places, where we seem to ourselves to
make ourselves and grow, are the parts of the world to which we are
closest, the parts of which our knowledge is the most intimate and
complete. Why should we not take them at their facevalue? Why may
they not be the actual turning-places and growing-places which they
seem to be, of the world---why not the workshop of being, where we
catch fact in the making, so that nowhere may the world grow in any
other kind of way than this?

Irrational!\ we are told. How can new being come in local spots and
patches which add themselves or stay away at random, independently of
the rest? There must be a reason for our acts, and where in the last
resort can any reason be looked for save in the material pressure or
the logical compulsion of the total nature of the world? There can be
but one real agent of growth, or seeming growth, anywhere, and that
agent is the integral world itself. It may grow all-over, if growth
there be, but that single parts should grow {\it per se\/} is
irrational.

But if one talks of rationality---and of reasons for things, and
insists that they can't just come in spots, what {\it kind\/} of a
reason can there ultimately be why anything should come at all?
\end{quotation}
then more modernly \cite{Wheeler82},
\begin{quotation}\small
Each elementary quantum phenomenon is an elementary act of ``fact
creation.'' That is incontestable. But is that the only mechanism
needed to create all that is? Is what took place at the big bang the
consequence of billions upon billions of these elementary processes,
these elementary ``acts of observer-participancy,'' these quantum
phenomena? Have we had the mechanism of creation before our eyes all
this time without recognizing the truth? That is the larger question
implicit in your comment [``Is the big bang here?''].
\end{quotation}
When cosmology hails from the inside, the world stands a chance of being anything it wants to be.

\section{The Future}

\begin{flushright}
\baselineskip=13pt
\parbox{2.8in}{\baselineskip=13pt\small
It is difficult to escape asking a challenging question. Is the
entirety of existence, rather than being built on particles or fields
of force or multidimensional geometry, built upon billions upon
billions of elementary quantum phenomena, those elementary acts of
``observer-participancy,'' those most ethereal of all the entities
that have been forced upon us by the progress of science?}
\medskip\\
\small --- John Archibald Wheeler
\end{flushright}

There is so much still to do with the physics of QBism; this article gives no hint.  Just one example:  The technical problems with SICs are manifest.  For instance, there must be a reason a proof of their existence has been so recalcitrant.  An optimist would say it is because they reach so deeply into the core of what the quantum is telling us!  In any case, we do suspect that when we get the structure of SICs down pat, Eq.~(\ref{ScoobyDoo}), though already so essential to QBism's distillation of quantum theory's message, will seem like child's play in comparison to the vistas the further knowledge will open up.

But the technical also complements and motivates the conceptual.  So far we have only given the faintest hint of how QBism should be mounted onto a larger empiricism.  It will be noticed that QBism has been quite generous in treating agents as physical objects when needed.  ``I contemplate you as an agent when discussing your experience, but I contemplate you as a physical system before me when discussing my own.''  Our solution to ``Wigner's friend'' is the great example of this.  Precisely because of this, however, QBism knows that its story cannot end as a story of gambling agents---that is only where it starts.  Agency, for sure, is not a derivable concept as the reductionists and vicious abstractionists would have it, but QBism, like all of science, should strive for a Copernican principle whenever possible.  We have learned so far from quantum theory that before an agent the world is really malleable and ready through their intercourse to give birth.  Why would it not be so for every two parts of the world?  And this newly defined valence, quantum dimension, might it not be a measure of a system's potential for creation when it comes into relationship with those other parts?

It is a large research program whose outline is just taking shape.  It hints of a world, a pluriverse, that consists of an all-pervading ``pure experience,'' as William James called it.\footnote{Aside from James's originals \cite{James96a,James96b}, further reading on this concept and related subjects can be found in Refs.~\cite{Lamberth99,Taylor96,Wild69,Gieser05,Russell08,Banks03,Heidelberger04}.}  Expanding this notion, making it technical, and letting its insights tinker with spacetime itself is the better part of future work.  Quantum states, QBism declares, are not the stuff of the world, but quantum {\it measurement\/} might be.  Might a one-day future Shakespeare write with honesty,

\begin{flushright}
\baselineskip=13pt
\parbox{3.4in}{
\begin{verse}\small
Our revels are now ended.  These our actors, \\
As I foretold you, were all spirits and \\
Are melted into air, into thin air \smallskip \ldots \\
We are such stuff as \\
\hspace*{0.5cm} quantum measurement is made on.
\end{verse}}
\end{flushright}

\section{Acknowledgments}

{\small\baselineskip=13pt Beside my direct collaborators D.~M. Appleby, H. Barnum, H.~C. von Baeyer, C.~M. Caves, and R. Schack, to whom I am always indebted, I thank my (sympathetic, agnostic, skeptical, whichever fits) friends \v{C}. Brukner, E.~G. Cavalcanti, C. Ferrie, R. Healey, N.~D. Mermin, W.~C. Myrvold, M. Schlosshauer, J.~E. Sipe, R.~W. Spekkens, C.~G. Timpson, and H.~M. Wiseman for supplying critical tension at the times I needed it, and especially L. Hardy for asking the right question in his inevitably civilized way---ways of asking questions can make all the difference.  Finally, not a joke, I thank S. Lloyd for demonstrating his shoe in a lecture many years ago; that shoe has left its mark in this essay.  Some of this essay was written with the kind hospitality of IQOQI, University of Vienna.  Work reported in Section \ref{SeekingSICs} was supported in part by the U.~S. Office of Naval Research (Grant No.\ N00014-09-1-0247).}

\end{document}